# Electric-Field-Controlled Antiferromagnetic Spintronic Devices

Han Yan, Zexin Feng, Peixin Qin, Xiaorong Zhou, Huixin Guo, Xiaoning Wang, Hongyu Chen, Xin Zhang, Haojiang Wu, Chengbao Jiang, Zhiqi Liu*

H. Yan, Z. Feng, P. Qin, X. Zhou, H. Guo, X. Wang, H. Chen, X. Zhang, H. Wu, Prof. C. Jiang, Prof. Z. Liu
School of Materials Science and Engineering, Beihang University, Beijing 100191, China
Email: zhiqi@buaa.edu.cn

**In recent years, the field of antiferromagnetic spintronics has been substantially advanced. Electric-field control is a promising approach to achieving ultra-low power spintronic devices via suppressing Joule heating. In this article, cutting-edge research, including electric-field modulation of antiferromagnetic spintronic devices using strain, ionic liquids, dielectric materials, and electrochemical ionic migration, are comprehensively reviewed. Various emergent topics such as the Néel spin-orbit torque, chiral spintronics, topological antiferromagnetic spintronics, anisotropic magnetoresistance, memory devices, two-dimensional magnetism, and magneto-ionic modulation with respect to antiferromagnets are examined. In conclusion, we envision the possibility of realizing high-quality room-temperature antiferromagnetic tunnel junctions, antiferromagnetic spin logic devices, and artificial antiferromagnetic neurons. It is expected that this work provides an appropriate and forward-looking perspective that will promote the rapid development of this field.**

## 1. Introduction

Antiferromagnetism is the manifestation of a magnetic order that has two or more magnetic sublattices aligned in such a manner that the total moment is zero. The antiferromagnetic order vanishes above the Néel temperature, named after Louis Néel, who first discussed this type of magnetism. In the Nobel lecture of Néel in 1970, he mentioned that antiferromagnets are extremely interesting from a theoretical viewpoint, but do not seem to have any applications.[1] Indeed, decades after they were discovered, antiferromagnets were utilized as auxiliary materials such as exchange-bias pinning layers in spin-value structures of magnetic recording devices. However, this is changing with the increased interest of antiferromagnetic spintronics in recent years.[2-42]

Compared to ferromagnets, antiferromagnetic materials have faster spin dynamics and low sensitivity to stray magnetic fields, which makes them ideal for spintronic memory devices. The stability of antiferromagnetic spins under external magnetic fields is related to the spin-flop field , where $H_E$ is the antiferromagnetic exchange field, and $H_A$ is the anisotropic field. Usually, the $H_{SF}$ of an antiferromagnet can reach a few tens to more than a hundred Teslas and is much higher than the $H_A$ that represents the steadiness of ferromagnetic moments. On the other hand, owing to the spin canting, the zero-field antiferromagnetic resonance frequency $\omega_{AFM}$ can be estimated by ,[39] where γ is the gyromagnetic ratio. It is, therefore, closely related to the Néel temperature and can reach THz. This indicates that the spins of an antiferromagnet can be switched on a timescale of a picosecond, which is three orders of magnitude faster than that of ferromagnets (~ns).

In 2014, Železný *et al*. theoretically proposed[43], and later in 2016, the same group creatively realized the room-temperature electrical switching of the antiferromagnetic spin axis in an antiferromagnetic CuMnAs thin film (**Figure 1**a).[14] An electrical current injected into the CuMnAs crystal with a central cross structure as shown in Figure 1b generates a spin polarization, which serves as an effective staggered magnetic field. The antiferromagnetic order parameter is thus rotated by 90º and tends to align to the orientation of the staggered field, which is perpendicular to the applied current. This switching mechanism is called Néel spin-orbit torque, as its process resembles the spin-orbit torque switching mechanism in ferromagnets. It yields two non-volatile resistance states (Figure 1c) based on the anisotropic magnetoresistance effect related to the anisotropy of the density of states and spin-orbit coupling. Furthermore, the applied current density is ~$4\times10^6$ A·cm$^{-2}$ and is much lower than that in ferromagnets (~$10^8$ A·cm$^{-2}$),[44] showing the relatively low energy consumption of the electrically controlled antiferromagnetic memory. In addition, the resistance states are stable under the influence of a large external magnetic field of 5 T (Figure 1d), which illustrates the insensitivity to external magnetic fields of the antiferromagnetic memory. This work achieved the first antiferromagnetic memory manipulated using electric current at room temperature, which largely promotes the progression of antiferromagnetic spintronic devices.

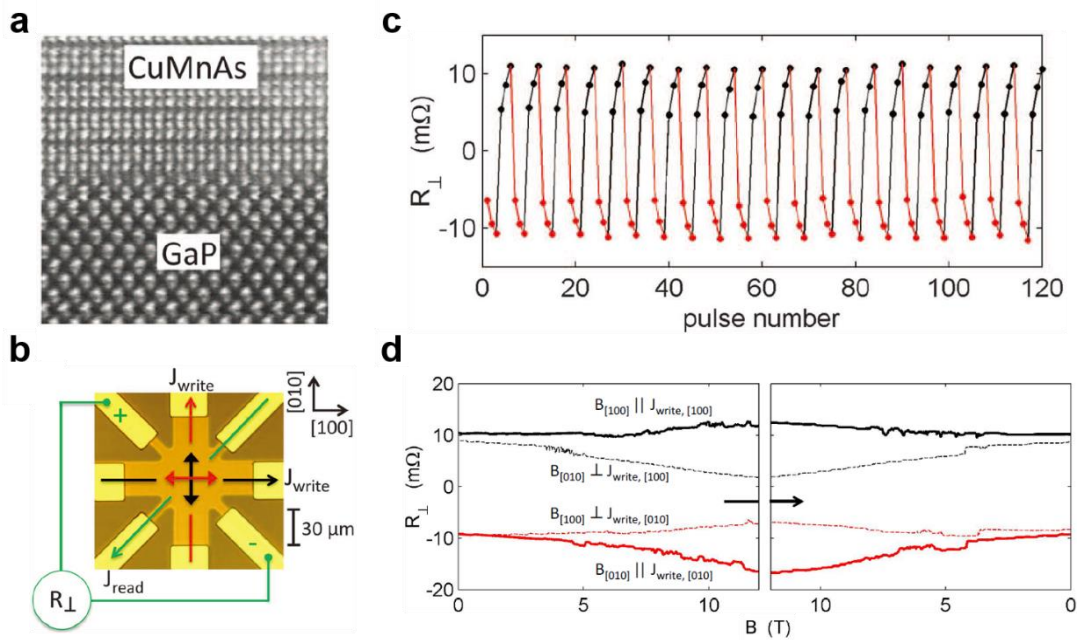

**Figure 1.** a) Scanning transmission electron microscopy image of the CuMnAs antiferromagnetic thin film growth on a GaP substrate. b) Microscope picture of the device and schematic of the measurement configuration. c) Manipulation of the transverse resistance by applied current pulses. d) Magnetic-field-dependent transverse resistance of the device.[14] Copyright 2016, American Association for the Advancement of Science.

Two years later, the THz electrical writing speed in the antiferromagnetic memory device based on CuMnAs was experimentally demonstrated by the same group.[45] They fabricated a device consisting of a GaAs substrate, Au contact pads, and a CuMnAs antiferromagnetic thin film with a central cross structure (**Figure 2**a and b). An incident THz electric field with different linear polarizations (Figure 2c) was used instead of an electric ac current to generate ultrashort current pulses (~1 ps) (Figure 2d), which switches the antiferromagnetic moments and achieves the information writing with a THz speed. The current density corresponding to the incident THz electric field is ~$2.7 \times 10^9$ A·cm$^{-2}$; however, the writing energy for the THz writing speed remains low due to the short duration time. Thus, this antiferromagnetic device exhibits the advantage of the THz writing speed with low energy dissipation.

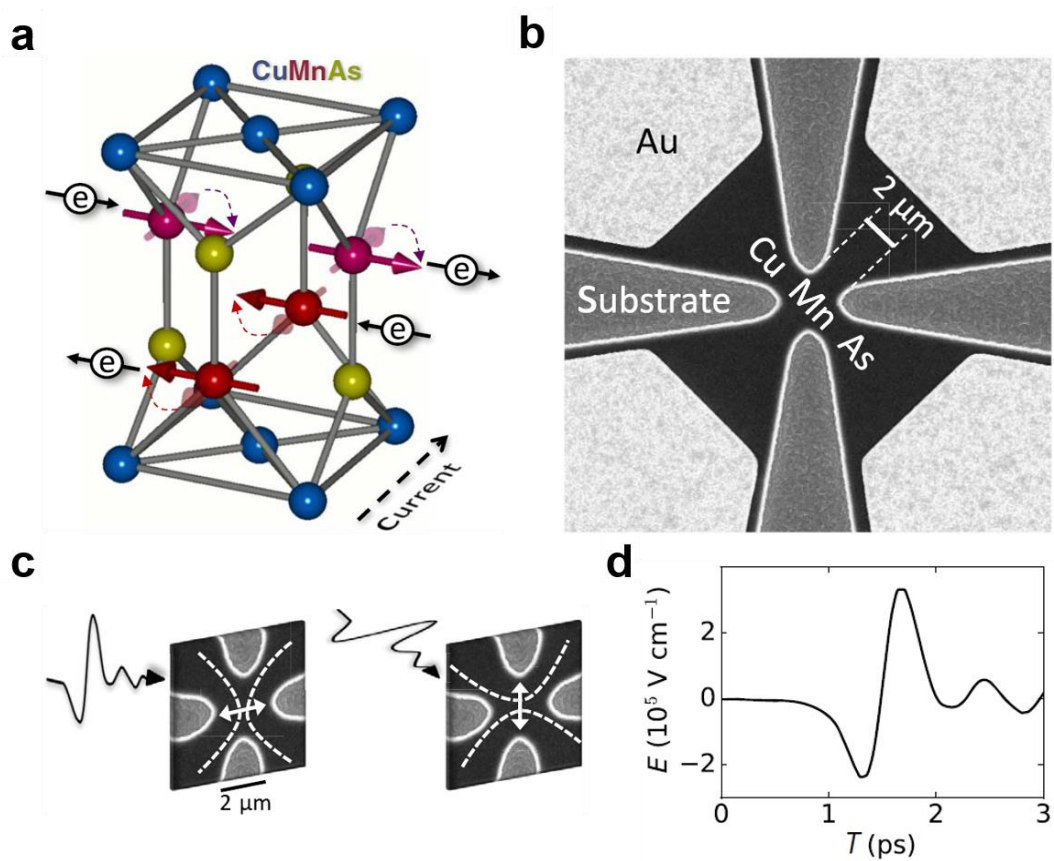

**Figure 2.** a) Illustrative diagram of the magnetic structure and the current-induced switching mechanism of the antiferromagnetic CuMnAs. b) Electron microscopy picture of the device. c) Sketches of the writing operation by incident THz electric field. d) Waveform of the applied picosecond writing pulses.[45] Copyright 2018, American Association for the Advancement of Science.

In addition to the resistance modulation of collinear antiferromagnets, the electrical switching of noncollinear antiferromagnetic memory devices was achieved as well. Hajiri *et al.* reported the switching of the magnetic order parameter of the noncollinear antiferromagnet Mn$_3$GaN through an electric current in an experiment in 2019 (**Figure 3**).[46] They found that the manipulation of the Hall resistance in the $Mn_3GaN$/Pt bilayer can be induced by a pulse current, which disappeared in the single $Mn_3GaN$ layer, and thus inferred the switching of noncollinear $Mn_3GaN$ antiferromagnet is caused by the spin-orbit torque. This work demonstrates the possibility of electrical switching of a noncollinear antiferromagnet using the spin-orbit torque. However, the exact physical mechanisms for the Hall resistance are not fully understood, as mentioned in the conclusion, and some other extrinsic effects such as Joule heating may have also contributed to the change of Hall resistance.

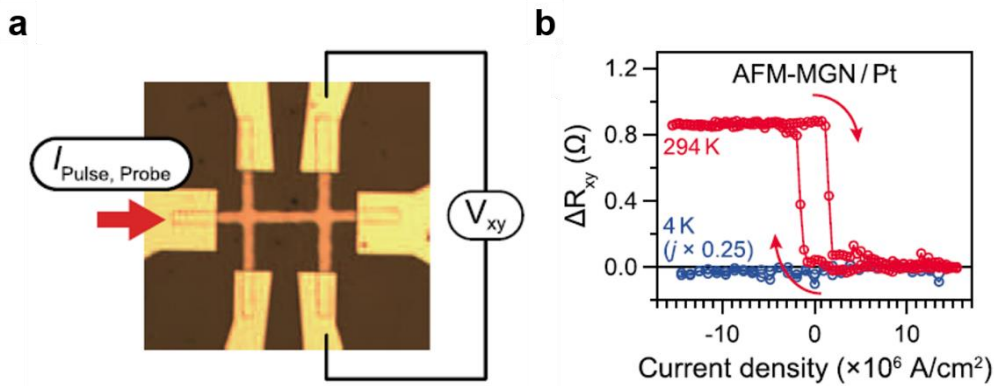

**Figure 3.** a) Optical microscopy image of antiferromagnetic $Mn_3GaN$/Pt bilayer and schematic of the experimental geometry. b) Dependence of the Hall resistance to the current density.[46] Copyright 2019, American Institute of Physics.

Despite the above intriguing resistance switching induced by external electrical currents, which have mainly contributed to electrical switching of Néel vectors in antiferromagnets, there are other reports that emphasize non-spin-orbit torque origins[47] or the absence of the electrical switching of Néel vectors[48] in antiferromagnets. For example, the magnetoelastic effect could be important[47], and Joule heating could be significant in current-switched antiferromagnetic spintronic devices.[48] A possible useful approach for determining whether the current switching effect is intrinsic or extrinsic is to use the spin-flop transition triggered by external magnetic fields, which for sure can manipulate the AFM spins, to examine the effect of spin structure change on Hall resistance and subsequently compare the current switching of Hall resistance.

Besides memory devices and mechanisms for electrical current switching of antiferromagnetic spins, antiferromagnets can be used as THz oscillators. In 2016, Cheng *et al.*[49] theoretically proposed a THz antiferromagnetic spin Hall nano-oscillator in an antiferromagnet/heavy-metal heterostructure using a dynamic feedback mechanism derived from the combined effect of the spin Hall effect and its reverse process, *i.e.*, the inverse spin Hall effect. The THz oscillation arises when the amplitude of the spontaneous motion induced by the spin-transfer torque is balanced to the magnetic damping. Subsequently, in 2017, Khymyn *et al.* theoretically suggested a THz antiferromagnetic oscillator driven by the spin current in a bilayer structure of Pt and NiO.[50] As shown in **Figure 4**, a spin current, which originates from the spin Hall effect of the Pt layer and is polarized along the hard axis of antiferromagnetic moments of NiO, is injected into the NiO layer, and then creates a spin-transfer torque. Consequently, the spin torque tilts the antiferromagnetic moments from their equilibrium opposite orientation and excites an effective field and leads to the rotation of the canted antiferromagnetic spin sublattices. Due to a weak easy-plane antiferromagnetic anisotropy, the rotation is non-uniform in time, thus causing an additional oscillation of THz oscillation frequency. Meanwhile, the THz spin-pumping induced by the oscillation returns to the Pt layer, generating a spin current that can be converted to a measurable electric current using the inverse spin Hall effect. Compared to the THz oscillator exploiting the feedback mechanism, this pattern has lower power consumption, because the former needs a large current density to overcome not only the magnetic damping but the antiferromagnetic hard axis anisotropy. Overall, these theoretical explorations of the THz antiferromagnetic oscillator provide promising paths for applications.

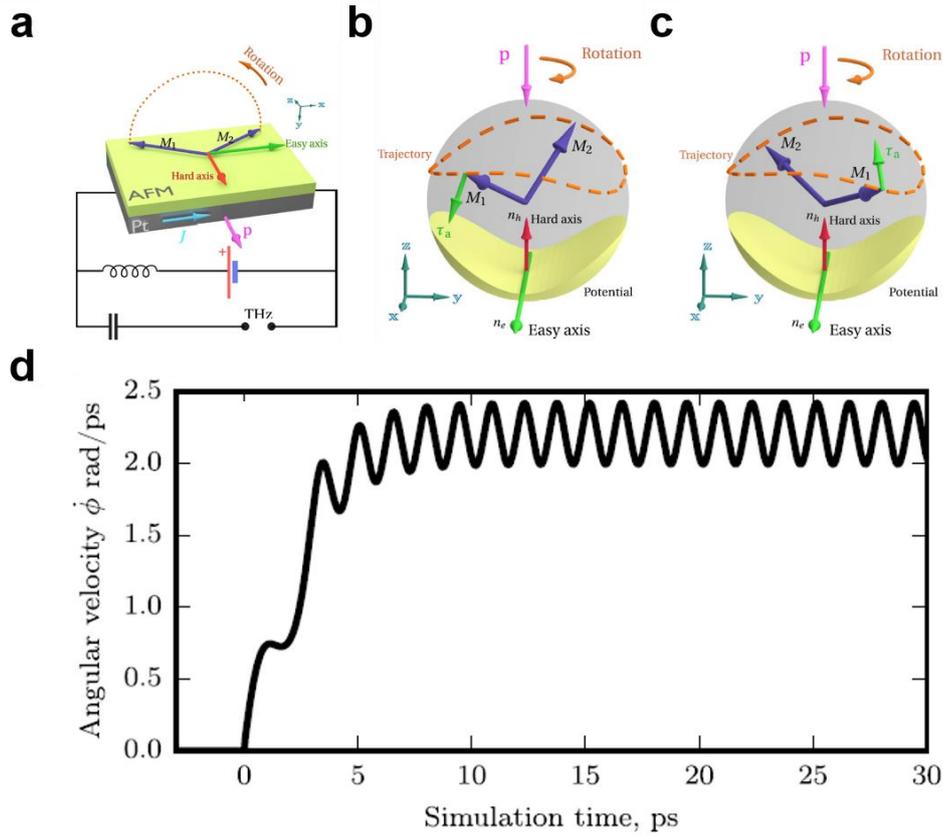

**Figure 4.** a) Schematic of the THz antiferromagnetic oscillator driven by spin current in a bilayer structure of Pt and NiO. b,c) Illustrative diagram of the rotation of the canted antiferromagnetic moments driven by the current-induced spin transfer torque. d) Temporal characteristics of the rotation of the antiferromagnetic moments in the oscillator.[50] Copyright 2017, Springer Nature-NPG.

Additionally, Kimata *et al.* creatively came up with a novel concept of the magnetic spin Hall effect and the magnetic inverse spin Hall effect in a noncollinear antiferromagnet $Mn_3Sn$.[51] In this study, they fabricated a device comprising of a NiFe/MgO bilayer pad and a Cu electrode pad based on a $Mn_3Sn$ single crystal substrate (**Figure 5**a). Then, they injected current into the $Mn_3Sn$ layer and applied a magnetic field that was rotated within the kagome plane of $Mn_3Sn$ and discovered a transverse voltage signal between the NiFe layer and the Cu electrode (Figure 5b). The resistance-magnetic field curve exhibited a rectangular hysteresis loop (Figure 5c), indicating that there was spin accumulation at the $Mn_3Sn$ surface. When the NiFe magnetization was switched following the rotation of the external magnetic field relative to the accumulated spins, which had a fixed polarization, the voltage signal across $Mn_3Sn$/NiFe interface was changed. More importantly, the hysteresis loop reversed when the triangular moments of $Mn_3Sn$ switched its orientation (Figure 5d), implying that the noncollinear magnetic order broke the time-reversal symmetry and thus induced the magnetic spin Hall effect. Similarly, the time-odd magnetic inverse spin Hall effect is also observed in $Mn_3Sn$. They theoretically explained the origin of the magnetic spin Hall effect, and its inverse process is the inter-band spin-density/electric-field response. The magnetic spin Hall and inverse spin Hall effects only appear in time-symmetry-broken systems, *i.e.*, magnetic systems, and must be odd under time reversal. This work discovers a new type of spin Hall effect, which is related to the magnetic orders in an antiferromagnetic system and extends the development of antiferromagnetic spintronics.

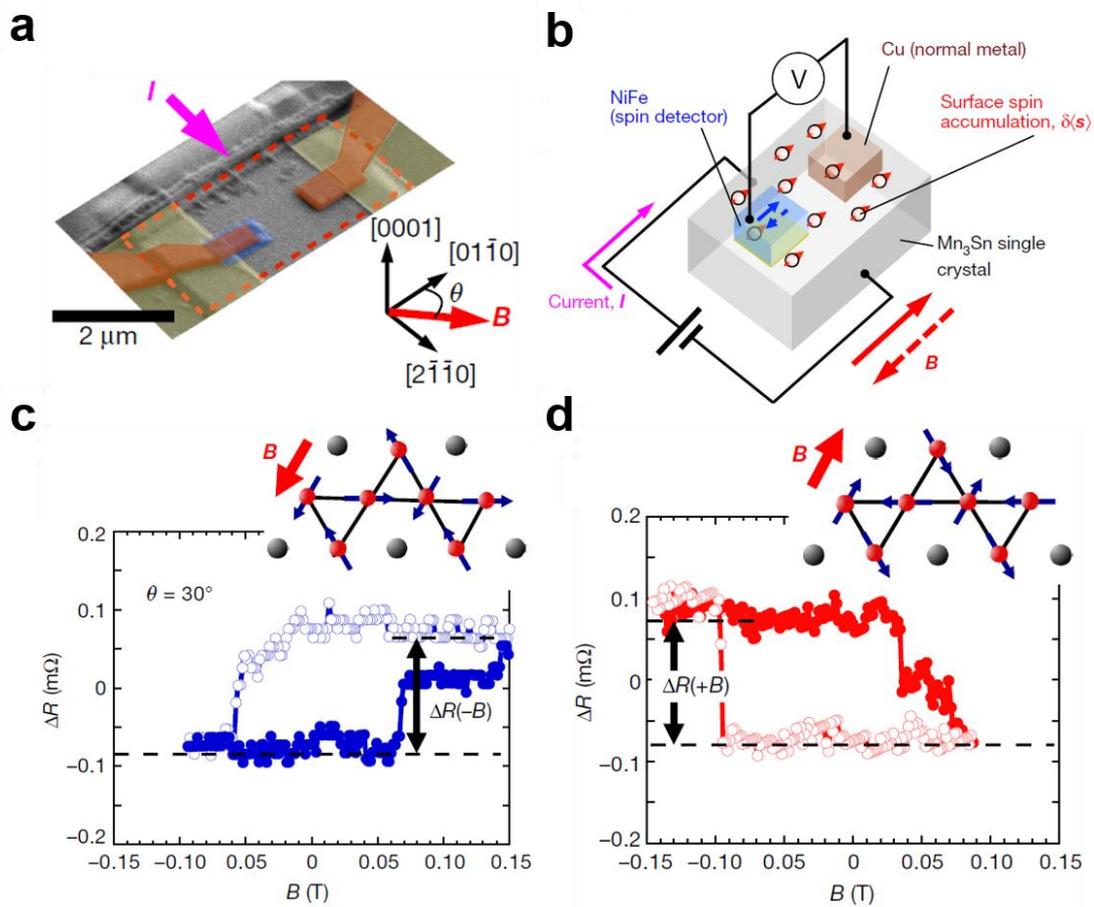

**Figure 5.** a) Scanning electron microscopy image of the device. b) Schematic of the device and the measurement scheme. c,d) Dependence of the resistance detected between NiFe and Cu on a scanning magnetic field. The orientation of the antiferromagnetic triangular moments of $Mn_3Sn$ in c) and d) is reversed as shown in the insets.[51] Copyright 2019, Springer Nature-NPG.

On account of these various theoretical and experimental studies, antiferromagnetic spintronics has been established and is in development. Nevertheless, the effective manipulation of the antiferromagnetic order parameter is quite challenging. The already successfully demonstrated approaches to controlling the antiferromagnetic spin axis include the exchange spring effect,[6] magnetic field assisted by thermal cycling across the antiferromagnetic order temperature,[7] spin-orbit torque,[14] and electric field[20,40] and so on. Particularly, the electric-field control of antiferromagnetism is attracting more and more attention owing to its unique advantage of low energy dissipation.[38,39]

## 2. Electric-field control of antiferromagnetic spintronic devices

### 2.1. Piezoelectric strain control

In the recent past, ferroelectric oxides have been widely used in the control of the magnetic order by forming ferromagnetic/ferroelectric or antiferromagnetic/ferroelectric composite heterostructures and has achieved effective modulation.[52-77] Ferroelectric oxides are effective insulating materials, possess spontaneous electric polarization that can be reversed with the application of a suitable electric field. Piezoelectric-strain modulation for functional thin films can be obtained using polarization switching of ferroelectric substrates. Utilizing piezoelectric strain induced by electric fields to realize antiferromagnetic devices with high energy efficiency can be achieved by integrating antiferromagnet materials onto ferroelectric substrates.

For example, $Mn_3Pt$, a cubic intermetallic alloy, has a first-order magnetic phase transition temperature of ~360 K. At low temperatures, $Mn_3Pt$ exhibits a non-collinear triangular antiferromagnetic state (**Figure 6**a) but converts into a collinear antiferromagnetic state above the transition temperature.[20] To compose an antiferromagnetic/ferroelectric heterostructure, high-quality $Mn_3Pt$ thin films were epitaxially grown on ferroelectric $BaTiO_3$ single-crystal substrates. Because of the intrinsic non-zero Berry curvature of the non-collinear antiferromagnetic state, an enhanced anomalous Hall effect exists below the transition temperature but is not there at the high-temperature collinear antiferromagnetic state (Figure 6b). In addition, along with the phase transition, a noticeable lattice expansion of ~0.8% occurs. Based on the sensitivity of the magnetic phases of $Mn_3Pt$ to its lattice structure, the transition temperature increases by ~25 K when a compressive piezoelectric strain of ~0.35% is applied (Figure 6c). Thus, at 360 K, the anomalous Hall effect can be switched on and off under electric fields of 4 kV·cm$^{-1}$ and 0 kV·cm$^{-1}$, respectively (Figure 6d). This work successfully demonstrated the piezoelectric-strain control of the anomalous Hall effect in antiferromagnet Mn$_3$Pt, which is of great significance for developing antiferromagnetic devices. Nevertheless, the optimal electric-field switching in this material with phase transition can only be achieved around its phase transition temperature. It is challenging to realize room-temperature antiferromagnetic spintronic devices unless certain elements are doped to Mn$_3$Pt to effectively reduce its phase transition temperature to 300 K.

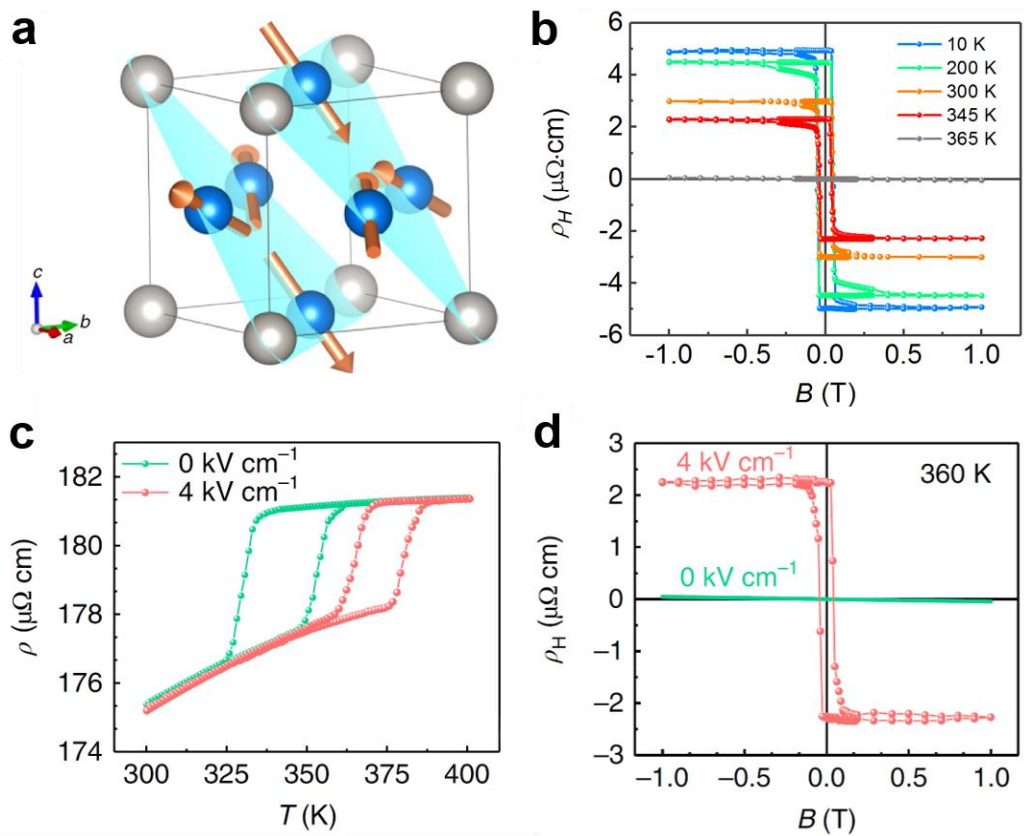

**Figure 6.** a) Crystal and non-collinear triangular magnetic structure of Mn$_3$Pt below 360 K. b) Anomalous Hall effect at different temperatures. c) ρ-T curves under an electric field of 0 and 4 kV·cm$^{-1}$. d) Hall effect under electric fields of 0 kV·cm$^{-1}$ and 4 kV·cm$^{-1}$.[20] Copyright 2018, Springer Nature-NPG.

Meanwhile, David *et al.* systematically researched the interface-induced strain effects on antiferromagnetic Mn$_3$NiN thin films by growing it on different substrates.[78,79] It was found that under biaxial or uniaxial strain, the symmetry of the antiperovskite structure (**Figure 7**a) could be destroyed, which leads to changes in the Néel temperature (Figure 7b) and magnetization (Figure 7c). For biaxial strain, the Néel temperature has an almost linear dependence on strain, and modulation of ~60 K by ±0.25% strain (Figure 7b). For uniaxial strain induced by a series of structural transitions of BaTiO$_3$, as temperature decreases, Mn$_3$NiN thin films transit from the ferrimagnetic state into the antiferromagnetic state and show a significant increase in magnetization at the orthorhombic to rhombohedral transition of BaTiO$_3$ at 187 K (Figure 7d), which implies a high sensitivity of the antiferromagnetic state to strain. Furthermore, David *et al.* mapped the temperature-biaxial strain magnetic phase diagram of Mn$_3$NiN according to the saturated Hall resistivity $\rho_{xy,sat}$ (Figure 7e and f), which can be useful for developing strain-controlled antiferromagnetic devices. Although this research shows promising results on strain modulated antiferromagnetic states using structural phase transitions of BaTiO$_3$, electric-field-controlled devices using piezoelectric strain have not been demonstrated.

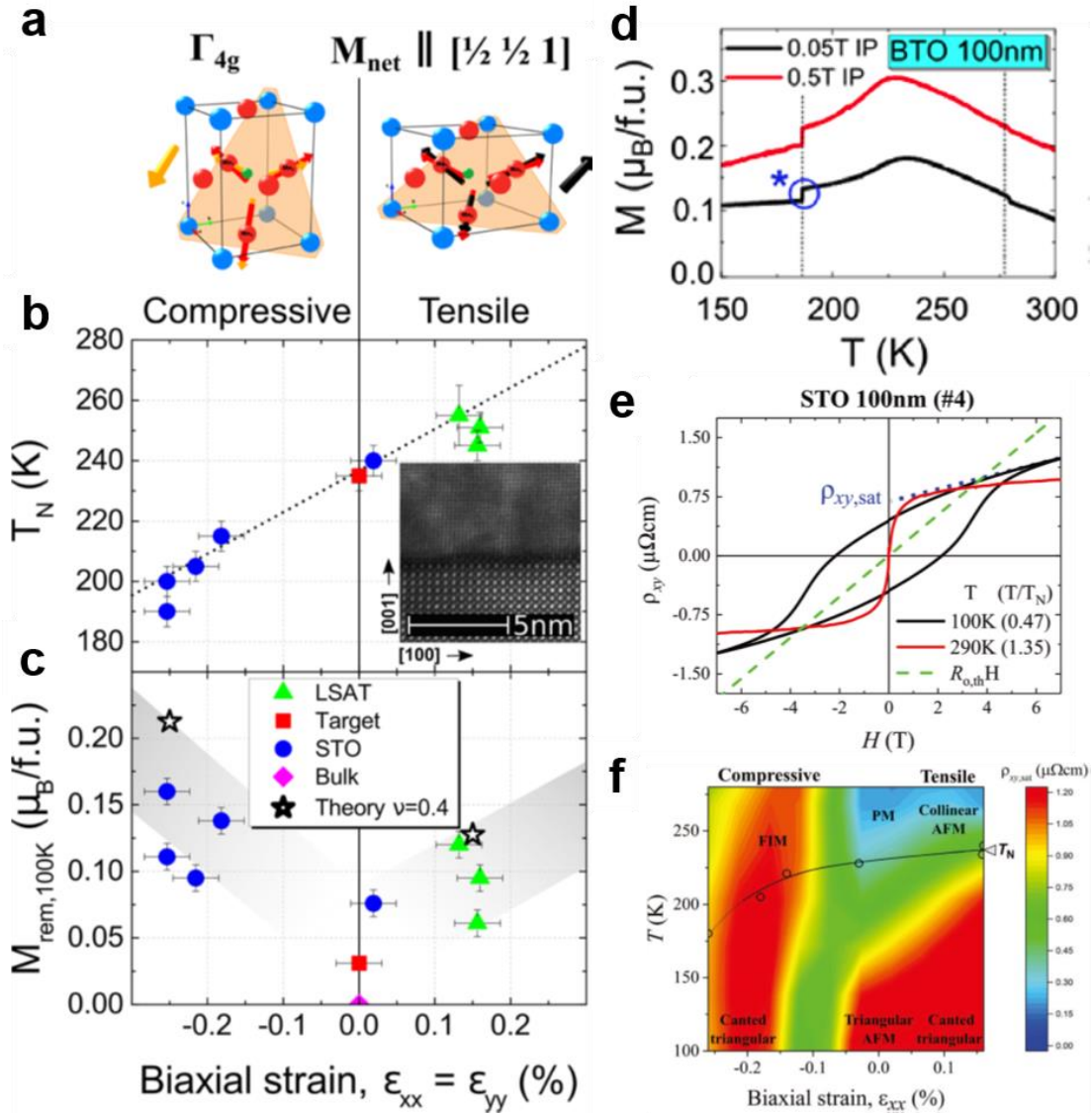

**Figure 7.** a) Crystal and magnetic structure of Mn$_3$NiN. Red arrows indicate the local moments of Mn. Yellow and black arrows indicate the magnetic moments under compressive and tensile strain, respectively. b,c) Biaxial strain dependence of Néel temperature and remnant magnetization. Inset is a cross-section transmission electron microscopy image of a Mn$_3$NiN film grown on SrTiO$_3$. d) Temperature-dependent magnetization of Mn$_3$NiN films grown on BaTiO$_3$.[78] Copyright 2018, American Chemical Society. e) Hall resistivity of Mn$_3$NiN films on SrTiO$_3$ as a function of magnetic field at 100 K and 290 K. The $\rho_{xy,sat}$ is estimated using a linear extrapolation taken from the high field region of $\rho_{xy}$ to the value at $H = 0$. f) Magnetic phase diagram as a function of biaxial strain and temperature.[79] Copyright 2019, Wiley-VCH.

The anomalous Hall effect in non-collinear antiferromagnets originates from the topological properties of Bloch bands and the resulting non-zero Berry curvature. In addition, there could be another kind of topological state in non-collinear antiferromagnets, Weyl fermions. For instance, Mn$_3$Sn has been the first non-collinear antiferromagnet discovered to exhibit the anomalous Hall effect.[12] It has been further theoretically predicted[80] and experimentally demonstrated[18] to a promising candidate of Weyl fermion systems. Accordingly, the piezoelectric control of its non-collinear spin structure could enable the realization of low-power and topologically-protected Weyl spintronic devices.

Our recent work[81] focuses on the integration of $Mn_3Sn$ thin films onto piezoelectric $0.72PbMg_{1/3}Nb_{2/3}O_3–0.28PbTiO_3$ (PMN-PT) single-crystal substrates. By optimizing growth conditions including temperature, sputtering power, and Ar pressure, we are able to realize the large anomalous Hall effect (**Figure 8**a and b) with low switching magnetic fields few mT fully comparable to the values of bulk $Mn_3Sn$.[12] More interestingly, the key features of the chiral anomaly for the Weyl fermion state are obtained as well, *i.e.*, negative parallel magnetoresistance (Figure 8c) and anisotropic magnetoresistance (Figure 8d). In addition, the insertion of a 100-nm-thick $LaAlO_3$ layer between the $Mn_3Sn$ film and the PMN-PT substrate was found to effectively suppress the crack[82] of the ferroelectric substrate upon cyclic electric fields, which ensures highly durable electric-field operation of the Weyl antiferromagnetic device (Figure 8e). It turns out that the anomalous Hall effect of $Mn_3Sn$ thin films is rather susceptible to piezoelectric strain, and a gate electric field of -3.6 kV·cm$^{-1}$ induces a ~190% modulation of the zero-field anomalous Hall resistance (Figure 8f). This reveals that the non-collinear spin structure of $Mn_3Sn$ is quite sensitive to strain and thus paves the way to Weyl antiferromagnetic spintronic devices. It is ideal for real applications if this modulation can be realized at room temperature by further improving the thin film quality.

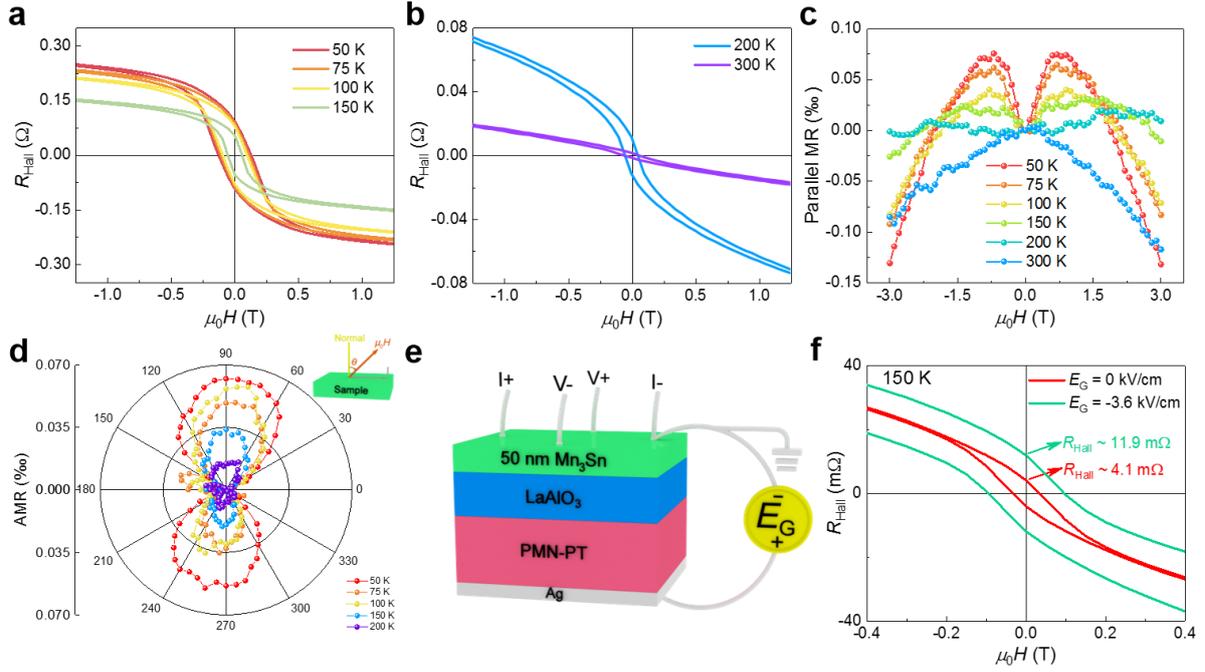

**Figure 8.** a) Low-temperature Hall effect of a 50-nm-thick $Mn_3Sn$/PMN-PT heterostructure deposited at a relatively low temperature of 150 °C. b) Hall effect of the same sample at 200 and 300 K. c) Magnetoresistance at different temperatures ranging from 50 to 300 K with the magnetic field applied parallel to the measuring current. d) Anisotropic magnetoresistance at various temperatures under 3 T. The angle of the magnetic field is defined relative to the normal of the sample surface. Inset: Schematic of the measurement configuration. e) Sketch of a $Mn_3Sn$ (50 nm)/$LaAlO_3$ (100 nm)/PMN-PT multiferroic heterostructure and the field-effect measurement geometry. f) Hall effect of the $Mn_3Sn$/$LaAlO_3$/PMN-PT heterostructure under gate electric fields $E_G$ = 0 and -3.6 kV·cm$^{-1}$. The electric field was exerted onto the heterostructure at 300 K and then kept during cooling down.[81] Copyright 2019, Elsevier.

Although the anomalous Hall effect in non-collinear antiferromagnets is of interest due to the exotic Berry curvature and in principle, the anomalous Hall resistance could be harnessed for information encoding in memory devices, the susceptibility of canted moments to magnetic fields in non-collinear antiferromagnets leads to the fact that the anomalous Hall resistance can be easily disturbed by external magnetic fields. Moreover, the electric-field-controlled anomalous Hall effect in non-collinear antiferromagnets has not been maturely realized at room temperature.[20,78,79,81] Instead, the piezoelectric-strain controlled longitudinal resistance[40] using the anisotropic magnetoresistance effect can well overcome these two bottleneck issues. By depositing textured MnPt films with preferred (101) and (001) orientations on ferroelectric PMN-PT substrates (**Figure 9**a), a continues modulation using piezoelectric strain at room temperature was achieved. The asymmetric butterfly shape of the electric-field-dependent resistance possesses two distinct states at zero electric field (Figure 9b).[40] More importantly, the high- and low- resistance state is non-volatile and can be switched by electric-field pulses at zero field, 9 and at 14 T (Figure 9c), which demonstrates a strain-controlled antiferromagnetic memory device insensitive to magnetic fields. In particular, the 14 T magnetic field is rather strong for real applications and is almost 9 times the surface magnetic field of the strong permanent magnet NdFeB. The main mechanism of this modulation is that under the biaxial compressive strain induced by the electric field, the antiferromagnetic axis of (101)-orientated grains are rotated towards the normal of the sample surface (Figure 9d). Owing to the strong antiferromagnetic coupling in MnPt, the electroresistance is found to be robust at a strong magnetic field of 60 T (Figure 9e). It would be ideal to realize nanoscale devices that are integrated on silicon by utilizing ferroelectric PMN-PT thin films. This would be an important research direction and the ultimate objective for strain-controlled antiferromagnetic memory devices.

Based on this work, Park *et. al.* carried out a theoretical study of strain-controlled spin axes of the antiferromagnetic MnX (X = Ir, Rh, Ni, Pd & Pt) alloys.[83] It was found that a small amount of strain can rotate the antiferromagnetic spin axes by 90° for all the materials. Specifically, for MnPt, the spin axis rotates from out-of-plane to in-plane under tensile strain, while the spin axes of MnIr, MnNi, MnRh and MnPd rotate within the basal plane.

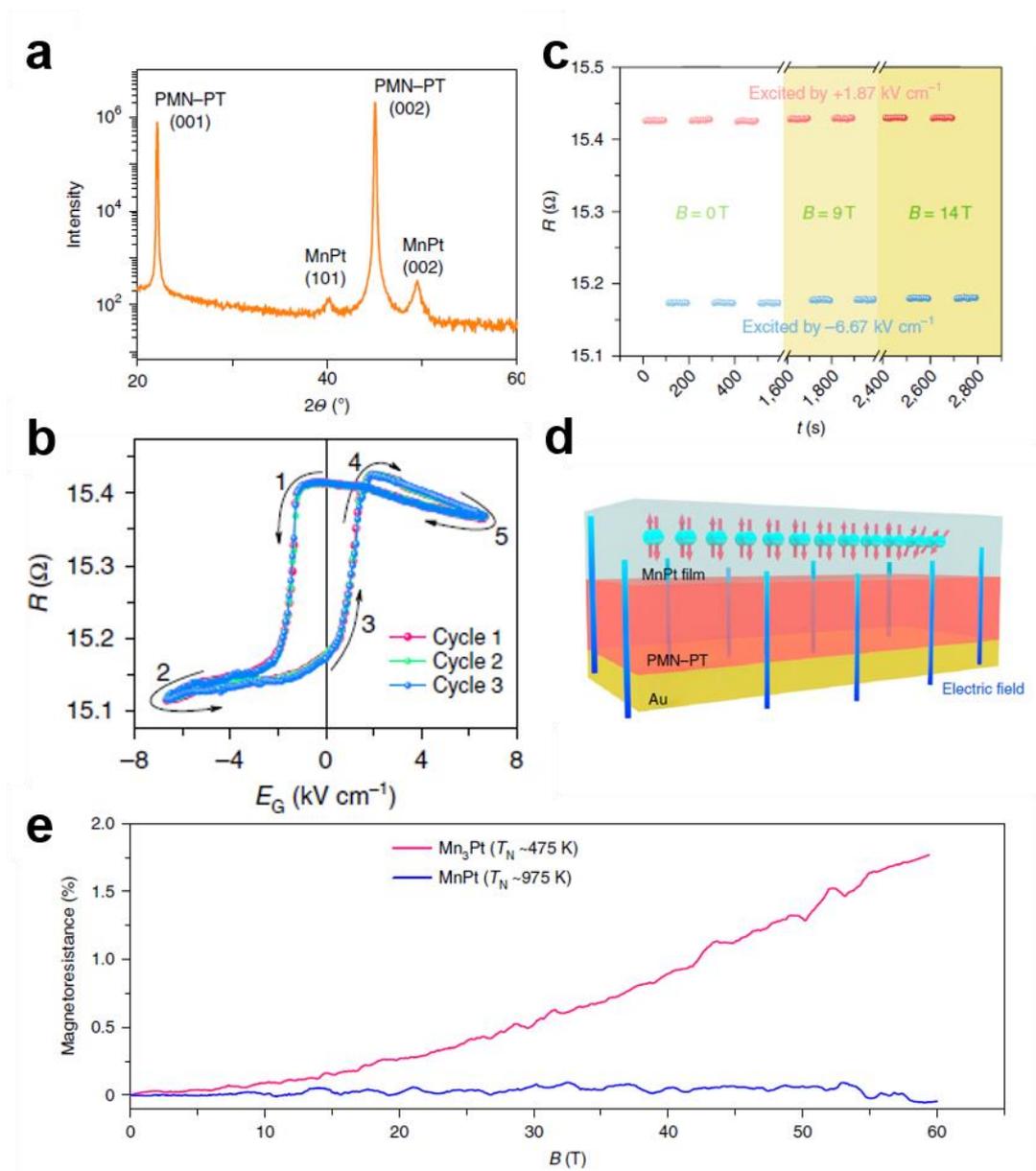

**Figure 9.** a) X-ray diffraction of a MnPt film grown on PMN-PT. b) Resistance of the MnPt film as a function of the electric field. c) The high-resistance state and the low-resistance state triggered by an electric field of +1.87 kV·cm$^{-1}$ and -6.67 kV·cm$^{-1}$ at room temperature under a magnetic field of 0, 9 and 14 T. d) Schematic of the spin axes distribution at the low-resistance state. e) Magnetoresistance of MnPt and Mn$_3$Pt films at a magnetic field up to 60 T.[40] Copyright 2019, Springer Nature-NPG.

Soon after the report of piezoelectric strain control of antiferromagnetic spins in MnPt thin films, Chen *et al*. have grown antiferromagnetic Mn$_2$Au thin films on (011)-oriented PMN-PT crystals and investigated the effect of piezoelectric strain on the uniaxial magnetic anisotropy of Mn$_2$Au.[84] Utilizing the small in-plane anisotropy of Mn$_2$Au, they found that the piezoelectric strain triggered by a positive electric field of +4 kV·cm$^{-1}$ and a negative electric field of -2 kV·cm$^{-1}$ could switch the easy axis from [100] to [0$\bar{1}$1] back and forth (**Figure 10**a).

Further studies showed the asymmetric Néel spin-orbit torque, and the corresponding antiferromagnetic ratchet also can be reversed by electric fields (Figure 10b). In general, the current-induced spin-orbit torque switching in antiferromagnets can obtain a resistance change of the order of ~10 mΩ, which is still small for real applications. Furthermore, in this work, the resistance difference between high- and low-resistance states is not very stable upon electrical cycling, which needs to be largely improved.

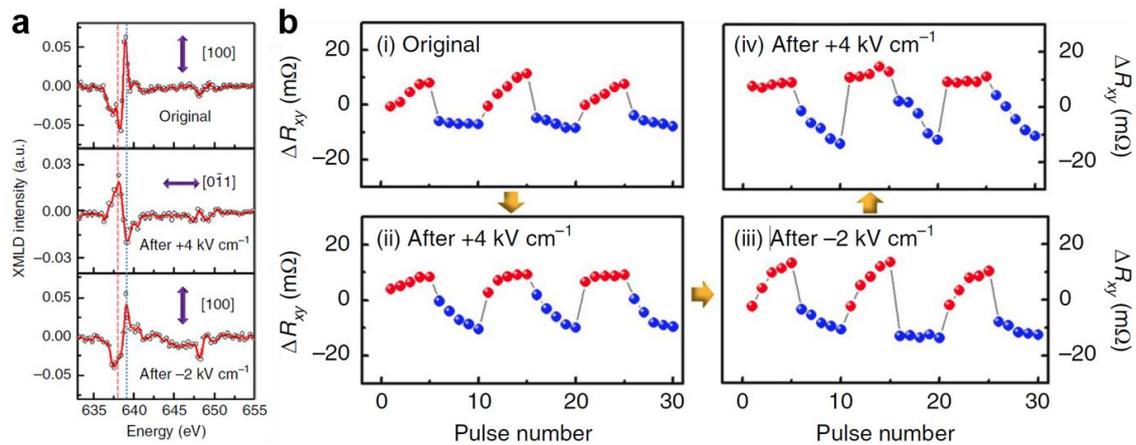

**Figure 10.** a) Mn *L*-edge XMLD signals in the original state and after excitation by an electric field of +4 kV·cm$^{-1}$ and -2 kV·cm$^{-1}$. b) Transverse resistance variation of the Mn$_2$Au device as a function of the number of current pulses.[84] Copyright 2019, Springer Nature-NPG.

Basically, although the absence of any stray field in antiferromagnets in sharp contrast to ferromagnets makes the magnetic field steering of antiferromagnetic spins rather challenging, the strain induces an addition magnetoelastic energy that originates from spin-orbit coupling and is equivalent to any materials with long-range magnetic ordering. From this perspective, the strain approach should be universal to engineer the anisotropy of magnetic materials. This line of thought also leads to the recent proposal of the new concept – antiferromagnetic piezospintronics.[39,40]

**2.2. Ionic Liquid modulation**

With the continuous development of condensed matter physics and spintronics, new mechanisms for modulation antiferromagnetic spins have emerged. For example, electric-field-controlled ionic liquid gating has drawn much attention and become a desirable method of manipulating antiferromagnetic materials. Compared with conventional ferroelectric strain, ionic liquid gating could also have the same advantages, such as fast response and low energy loss. However, the ferroelectric gating can maximally modulate a carrier density of ~$10^{14}$ cm$^{-2}$ due to limited polarization.[85] Ionic liquid gating contains a more powerful impact and penetrates a deeper thickness of several nanometers rather than a thickness of 1-2 atomic layers and can yield a significant carrier density modulation of ~$10^{15}$ cm$^{-2}$.[86] Moreover, ionic liquid gating needs a low voltage of a few tenth or few Volts. The large carrier modulation could affect the magnetic properties of antiferromagnetic materials to achieve antiferromagnetic spintronic devices. In addition, ionic liquid gating can induce ionic migration in oxides, such as oxygen or hydrogen,[87,88] which could achieve ionic modulation of magnetic properties, particularly in oxides.

In 2015, Wang *et al.*[89] reported that the exchange spring could be controlled in antiferromagnetic MnIr thin films by an ionic liquid. They patterned MnIr of different thicknesses and [Co/Pt] multilayers into Hall devices with a 2-nm-thick HfO$_2$ capping to prevent a chemical reaction between the metallic MnIr and the ionic liquid (**Figure 11**a). When a negative field was applied, the Hall measurements showed that the hysteresis loop shifted towards a negative position, and the exchange bias effect was enhanced. Upon applying a positive field, the changes were reversed (Figure 11b and c). They suggested that the origin of the phenomenon was the change of carriers in MnIr, and as a result, the intrinsic electronic structure and magnetic moment of Mn varied. Besides, the modulation was reversible, and the ordinary resistance varied ~0.2 % at 10 K and could be controlled by the electric field as well (Figure 11d). This work provides a novel means to manipulate exchange spring in antiferromagnetic materials for realizing low-power antiferromagnetic memory devices. Nevertheless, the potential drawback for this type of exchange biased multilayer systems is that the antiferromagnetic spins in IrMn are rather sensitive to external magnetic perturbations because of the small coercivity field of the ferromagnetic CoPt and the exchange spring effect between the ultrathin IrMn and the ferromagnetic CoPt. In contrast, the ionic liquid modulation of a single antiferromagnetic component without a ferromagnetic layer, such as in the following works[90,95] would not have such an issue.

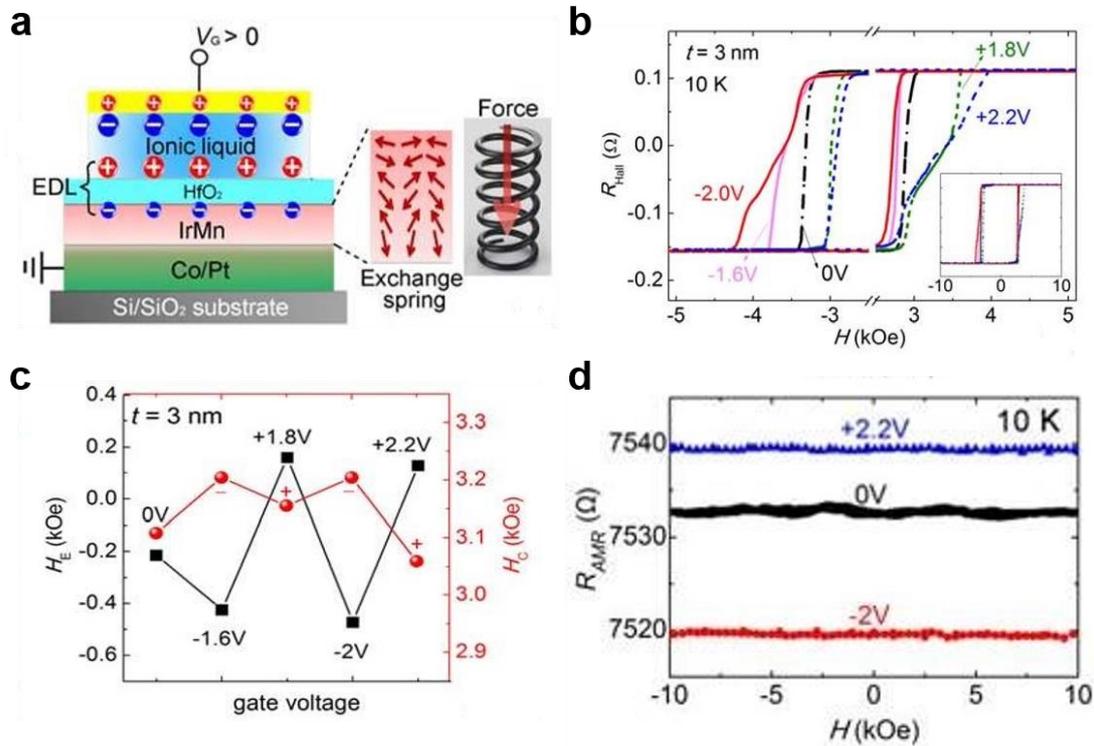

**Figure 11.** a) Schematic of charge distribution under positive gate voltage in the stack and the spin structure for MnIr exchange spring. b) Anomalous Hall effect curves for MnIr (3 nm)/[Co/Pt] multilayers with different applied field. c) The $H_E$ and $H_C$ extracted from anomalous Hall effect curves at 10 K. d) Magnetoresistance with vertical field up to 10 T induced by applied voltage.[89] Copyright 2015, Wiley-VCH.

In 2017, Lu et al.[90] creatively realized a tri-state phase transformation based on a selective dual-ion (oxygen and hydrogen) switch controlled by ionic liquid gating. The three different phases (the perovskite ferromagnetic metal $SrCoO_{3-\delta}$, the brownmillerite antiferromagnetic insulator $SrCoO_{2.5}$, a hitherto-unexplored and weakly ferromagnetic insulator $HSrCoO_{2.5}$) transform reversibly under the excitations of electric fields with associated electrochromic and magnetoelectric effects (**Figure 12**a and b). They presented that oxygen ions in the ionic liquid migrate to the $SrCoO_{2.5}$ layer, which then transforms to $SrCoO_{3-\delta}$ phase under a negative field, while under a positive field, hydrogen ions migrate to the $SrCoO_{2.5}$ layer and make the transformation to $HSrCoO_{2.5}$ occur (Figure 12c). In addition, the field-controlled multi-state resistance switch at room temperature (Figure 12d) could be of great significance in the application of random-access memory and this work paves the way to exploring the field-control of multi-state phase transition with rich functions. In addition, such an approach has been demonstrated to be effective for other materials as well.[91-94]

Typically, the electrostatic effect using ionic liquid gating is rather remarkable for modulating electrical and magnetic properties of thin films as it can change the carrier density by ~$10^{15}$ cm$^2$. However, the change of the voltage applied to the ionic liquid can only be feasible above the melting temperature of the ionic liquid instead of the low-temperature frozen state. In addition, the relaxation process for stable modulations takes a long time ~30 mins. These facts prevent it from being used in real high-speed, isothermal memory applications. Moreover, it is not compatible with the high-density integration technique due to complex liquid-solid interfaces.

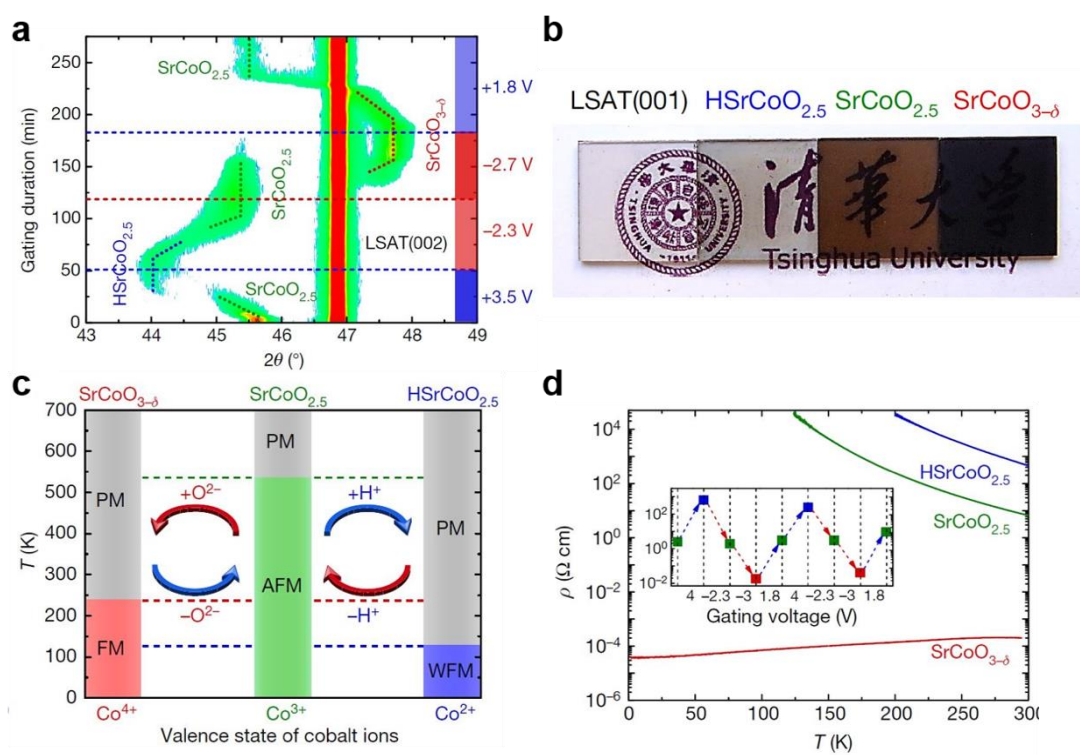

**Figure 12.** a) *In-situ* X-ray diffraction with varied electric fields. b) Electrochromic phenomenon. c) Schematic of tri-state reversible phase transformation due to the migration of hydrogen and oxygen ions in SrCoO$_{2.5}$. d) Temperature-dependent resistance and the inset exhibits a mode of tri-state resistive switch.[90] Copyright 2017, Springer Nature-NPG.

Almost at the same time, Zheng *et al.*[95] realized the ambipolar regulation of both electron and hole carriers in antiferromagnetic LaMnO$_3$ by ionic liquid gating. This work illustrated under the application of a positive (or negative) voltage, the induced electron doping (or hole doping) enables a ferromagnetic phase in an ultrathin 3-uc-thick LaMnO$_3$/SrTiO$_3$ heterostructure. When the applied voltage is ±3 V, and the magnetic field is 7 T, the anomalous Hall effect measurements reveal a symmetric feature between the hole and electron doping (**Figure 14**a). Although both electron and hole doping result in the colossal magnetoresistance effect, the induction of the colossal magnetoresistance effect by holes and electrons is asymmetric (Figure 14b and c), and this may be attributed to the different sensitivities of different types of carriers to exchange coupling (Figure 14d). Motivated by these experimental results, ambipolar spintronic devices might become a reality and provide a brand-new research field in the future, although the magnetic phase transition temperature of this manganite is far below 300 K, which hinders it from practical room-temperature device applications.

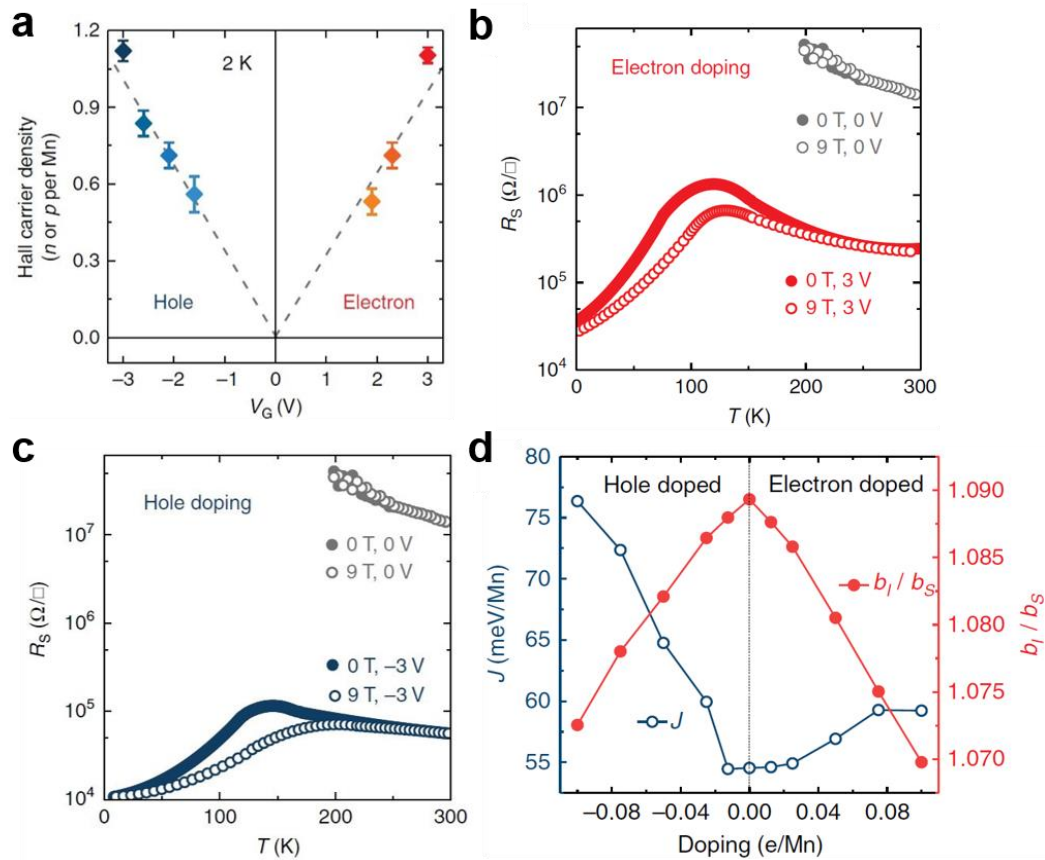

**Figure 14.** a) Hall carrier (electron or hole) density at 2 K in a 3 uc LaMnO$_3$ film. b,c) Temperature-dependent magnetoresistance in the electron-doped and hole-doped regions, respectively. d) Exchange coupling parameter $J$ and distortion parameter $b_l/b_s$ in two modes, where $b_l$ represents the longer bond between Mn$^{3+}$ ions and oxygen and $b_s$ is the shorter bond between Mn$^{3+}$ ions and oxygen.[95] Copyright 2018, Springer Nature-NPG.

Most of the antiferromagnetic oxides are poor conductors. As a result, the electrical signal read-out is relatively challenging unless the spin-wave approach, together with the inverse spin Hall effect is utilized, which typically needs a complicated multilayer structure with a heavy metal top layer such as Pt and the additional material layers for the generation of the spin-wave. Therefore, the recent ionic liquid modulation of the synthetic antiferromagnetic metal represents a promising advancement.[96]

Specifically, Yang *et al.*[96] demonstrated that the electric field on the base of the ionic liquid gating process could modulate Ruderman–Kittel–Kasuya–Yosida interaction in synthetic antiferromagnetic multilayers. They indicated that the electric field changes the Fermi level and correspondingly alters the strength of interlayer exchange coupling. In this work, the ferromagnetic and antiferromagnetic coupling of the synthetic antiferromagnets with the in-plane magnetic anisotropy (FeCoB/Ru/FeCoB) (**Figure 13**a and b) and perpendicular magnetic anisotropy ((Pt/Co)$_2$/Ru/(Co/Pt)$_2$) (Figure 13c and d) was controlled by a low voltage. Surprisingly, they reversed up to 80 % of perpendicular magnetic moments with a bias voltage of 2.5 V at room temperature (Figure 13e and f). Moreover, they also predicted the electric-field modulation of magnetization with zero magnetic field. Owing to the risk of leakage and contamination of the ionic liquid, they proposed that ionic liquid gel would be a better alternative.

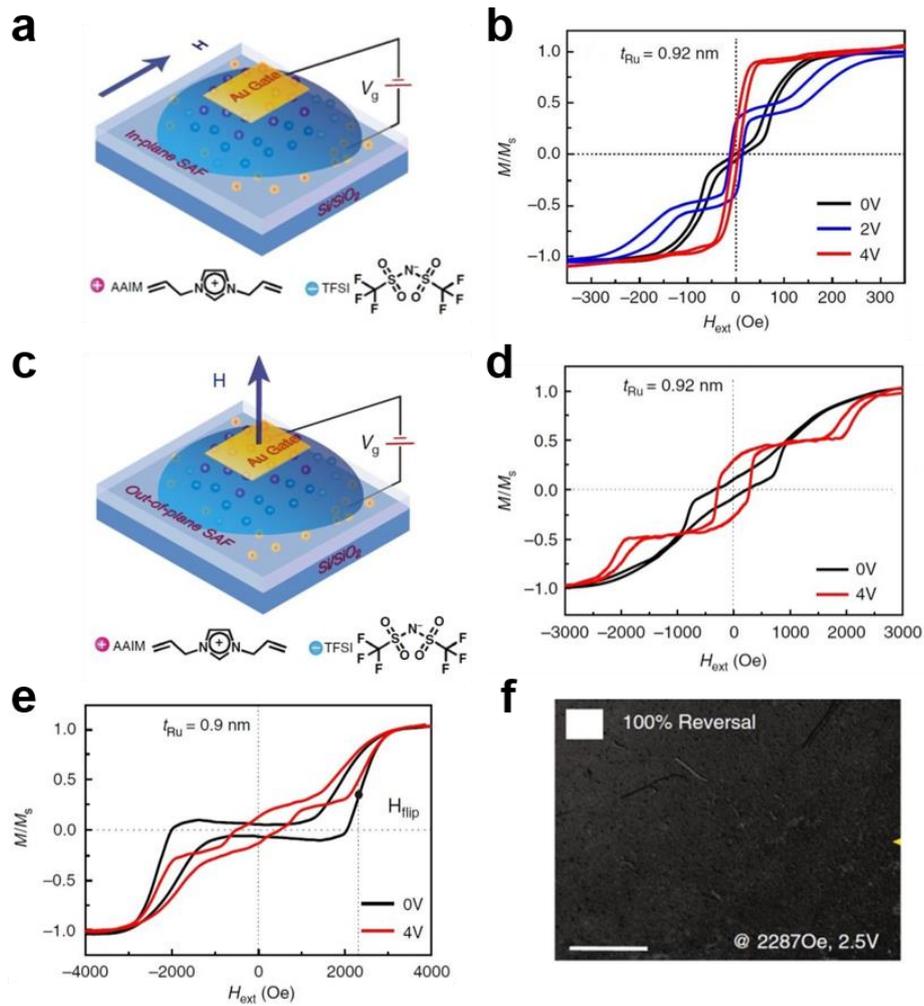

**Figure 13.** a,c) Schematics of in-plane (FeCoB/Ru/FeCoB) and out-of-plane ((Pt/Co)$_2$/Ru/(Co/Pt)$_2$) synthetic antiferromagnetic controlled by ionic liquid gating process. b,d) Ruderman–Kittel–Kasuya–Yosida interaction modulated by voltage when the thickness of Ru is 0.92 nm. e) *Ex-situ* magnetism measurement of out-of-plane synthetic antiferromagnet with $t_{Ru}$=0.9 nm. f) An image of domain switching at 2.5 V applied electric field.[96] Copyright 2018, Springer Nature-NPG.

## 2.3. Electrostatic carrier modulation by dielectric materials

Apart from the methods mentioned above, electrostatic modulation using dielectric materials is another way to modify the properties of materials, including electronic or magnetic features. Unlike chemical doping to tailor the carrier concentration of a material system, electrostatic modulation can reversibly and continuously alter the carrier density to cause the modification of the properties without changing the degree of the disorder.[97] The basic principle of the electrostatic modulation, in general, is applying an electric field to any material which attracts or dispels charges to form a layer where charges accumulate or deplete, and consequently changes the carrier density of the material to modify its electronic ground states.

The width of the accumulation and depletion layer is an essential parameter in field-effect experiments, which is called electrostatic screening length $\lambda_{el}$. In ordinary metals, $\lambda_{el}$ is very short, just about a part of the atomic radius,[98] where the electrostatic doping could hardly work. However, in low-carrier-density materials, such as correlated oxide insulators or semiconductors, the electrostatic doping could be effective.

Novel electronic phases occur in correlated oxides, while the carrier densities of these materials are at the order of $10^{19}$-$10^{22}$ carriers cm$^{-3}$.[98] Accurate tuning of their electrical and magnetic phases requires an accumulation and depletion level of the order of $10^{14}$ charges cm$^{-2}$. Here, we summarize a few studies that use electrostatic doping based on dielectric materials to modulate the electronic and magnetic properties in high-temperature superconductors and colossal magnetoresistive materials with an antiferromagnetic ground state in their parent materials.

In 1999, Ahn *et al.*[99] reported a non-volatile and reversible modulation of superconductivity transition temperature in the high-temperature copper oxide superconductor GdBa$_2$Cu$_3$O$_{7-x}$ (GBCO) by electrostatic doping. They fabricated Pb(Zr$_x$Ti$_{1-x}$)O$_3$(PZT)/GBCO heterostructures, in which the electric polarization of ferroelectric oxide PZT controls the Fermi level. At a low hole doping level, this material is an antiferromagnetic insulator. As the doping level is raised, the system becomes a superconductor. The phase diagram suggests that the doping level can solely control the superconductor-insulator transition, as well as the antiferromagnetic transition in the GBCO system.

In 2003, Hong *et al.*[100] presented a method to manipulate the magnetic transition temperature and magnetotransport properties by the electrostatic field effect. They combined a 4-nm-thick La$_{1-x}$Sr$_x$MnO$_{3-x}$ (LSMO) film with a 300-nm-thick ferroelectric oxide PZT film, which can supply a large polarization of 15-46 μC·cm$^{-2}$. At a low hole doping level, the system is at an antiferromagnetic insulating state. However, when the doping level $x$ is between 0.15 and 0.25 holes per unit cell, both magnetic and metal-insulator transitions occur. At this doping range, the colossal magnetoresistance effect happens near the Curie temperature. **Figure 15**a and b show the electrostatic modulation of the colossal magnetoresistance effect in the accumulation state, and depletion state of holes, respectively. The Curie temperature of the LMSO system was shifted by 35 K between the two different polarization states, and the magnetoresistance ratio was also changed, as depicted in Figure 15c.

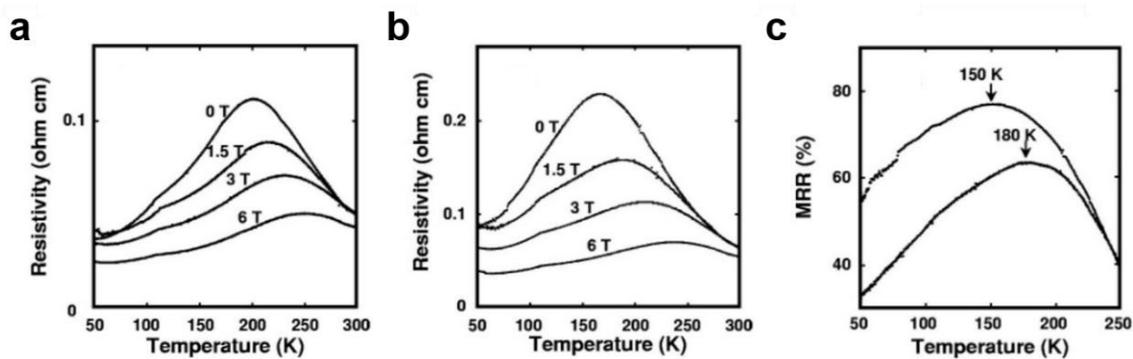

**Figure 15.** Resistivity as a function of temperature for a) the accumulation state and b) the depletion state of holes in LSMO at the magnetic field of 0, 1.5, 3 and 6 T. c) Magnetoresistance versus temperature at 6 T at both polarization states.[100] Copyright 2003, American Physical Society.

For strongly correlated oxide thin films, there is a general "dead layer" issue, *i.e.*, below certain thickness the originally metallic oxides become highly insulating and the originally magnetic oxides turn into non-magnetic materials due to the depletion of free carriers, defects, quantum confinement and other physical mechanisms in ultrathin films. Therefore, it is challenging to make the 4-nm-thick LSMO films[100] even thinner to further enhance the electrostatic modulation effect using dielectric materials. In contrast, the emerging layered magnetic two-dimensional materials have provided exciting opportunities regarding this aspect.

Recently, the discovery of two-dimensional van der Waals magnetic semiconductors[101,102] has attracted interests[103-105] owing to not only the novel interface phenomena in van der Waals heterostructures but also the possibility of electrostatic control of magnetism. Moreover, the inversion-symmetry-break-induced anomalous Hall effect has been discovered in non-magnetic two-dimensional layer materials as a result of the non-vanishing Berry curvature.[106-108] To date, a considerable amount of studies have achieved the electrostatic modulation of magnetic properties of ultrathin layer materials[109-112], including antiferromagnetic bilayer CrI$_3$.

Bilayer CrI$_3$ is a layered antiferromagnet that has a relatively low Néel temperature of ~45 K.[109] In bilayer CrI$_3$, spins within each layer are aligned out of plane ferromagnetically and then the two layers are coupled antiferromagnetically. At low temperatures, the antiferromagnetic-ferromagnetic metamagnetic transition can be induced by a low magnetic field of ~0.6-0.7 T,[109] which implies that only a subtle interlayer exchange interaction exists. This phenomenon reveals a fantastic opportunity to achieve a significant modulation of the antiferromagnetic order by even a slight carrier concentration change because the free electrons in semiconducting CrI$_3$ serve as vital mediators among itinerant moments.

In 2018, Jiang *et al*.[110] demonstrated an efficient control of antiferromagnetism of bilayer CrI$_3$ using electrostatic doping. They fabricated a dual-gate field-effect device in which CrI$_3$ was encapsulated in graphene and hexagonal boron nitride. Graphene is used to be electrodes, and hexagonal boron nitride is used as gate dielectrics. They utilized magnetic circular dichroism to probe the magnetization of bilayer CrI$_3$. **Figure 16**a shows that the spin-flip transition field increases significantly with the electron doping level. It finally drops to zero at a critical density of ~2.5×10$^{13}$ cm$^{-2}$, as given in Figure 16b. Consequently, they realized the electrostatic control of antiferromagnetism-ferromagnetism transition under zero magnetic field, purely by electrostatic carrier doping.

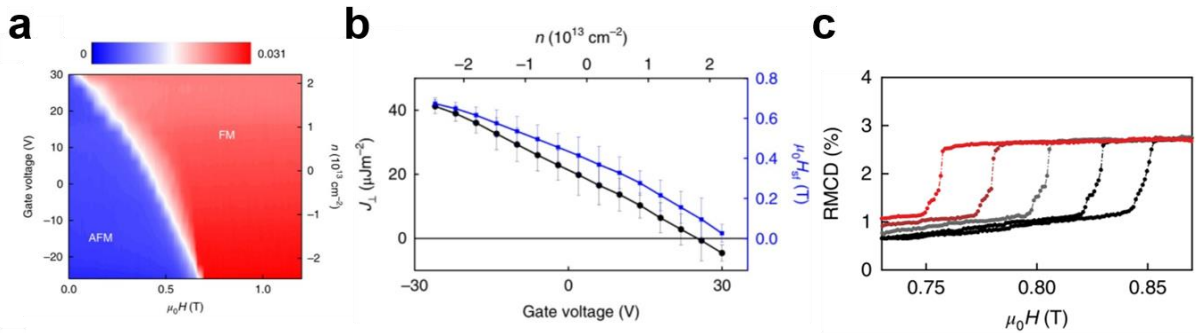

**Figure 16.** a) Schematic magnetic field-gate voltage magnetic phase diagram at 4 K. b) Gate voltage (bottom axis) and doping density *n* (top axis) dependence of the interlayer exchange constant $J_\perp$ (black,

right axis) and the spin-flip transition field (blue, left axis).[110] Copyright 2018, Springer Nature-NPG. c) Reflectance magneto-circular dichroism signal of a bilayer CrI$_3$ film versus applied magnetic field at a carrier density changing from 0(black) to 4.4×10$^{-12}$ cm$^{-2}$(red) at a step of 1.1×10$^{-12}$ cm$^{-2}$.[109] Copyright 2018, Springer Nature-NPG.

Almost at the same time, in 2018, Huang *et al.*[109] also used the electrostatic method to control the antiferromagnetism in bilayer CrI$_3$. Reflectance magneto-circular dichroism microscopy was used for probing magnetization. As shown in Figure 16c, at 15 K, the spin-flip field can be tuned by electrostatic doping in an applied magnetic field. This phenomenon indicates that the electrostatic doping can primarily drive the antiferromagnetism-ferromagnetism transition in bilayer CrI$_3$.

Generally, the long-range magnetic ordering in two-dimensional materials exhibits low ordering temperature probably because of sharp quantum spin fluctuations. Therefore, most of the spintronic devices built on these materials have been demonstrated at low temperatures. A means for overcoming this bottleneck could be the modulation of electron-electron correlation using tuning carrier density. For example, in the recent work by Deng *et al.*[112] the room-temperature magnetic ordering is obtained by electrostatic doping using ionic liquid gating. In addition, the antisite defects in intermetallic compounds may be helpful for stabilizing a higher spin ordering temperature.[113]

## 2.4. Electrochemical control of antiferromagnetism

Lastly, electrochemical modulation using ionic migration could be another way to tailoring antiferromagnetism. In 2014, Bi *et al.*[114] demonstrated that magnetic properties, including the saturation magnetization and anisotropy field of ultrathin Co films adjacent to Gd$_2$O$_3$ gate oxides, could be significantly modulated by electric fields in a non-volatile way. They patterned the samples which comprise of a Si/SiO$_2$/Pt(4 nm)/Co(0.7 nm)/Gd$_2$O$_3$(80 nm)/Ta(5 nm)/Ru(100 nm) multilayer structure into Hall bars for transport measurements (**Figure 17**a). The electric fields were applied at temperatures ranging from 200 ˚C to 260 ˚C, and then the anomalous Hall effect measurements were implemented at room temperature with zero electric field. Firstly, they measured the anomalous Hall resistance of the sample after staying at 200 ˚C for 10 min without any application of the electric field. Subsequently, the anomalous Hall effect vanished upon applying a small electric field of -625 kV/cm. Surprisingly, following a positive field application for 13 min at 200 ˚C, the $R_H$–$H_Z$ curve was nearly recovered to the initial shape (Figure 17b). Considering that $Gd_2O_3$ is an ionic conductor with high $O^{2-}$ mobility, $O^{2-}$ can be driven towards the Co layer under a negative electric field, consequently oxidizing the ultrathin ferromagnetic Co film into antiferromagnetic CoO. Reversibly, the positive voltage could drive the $O^{2-}$ in Co/$CoO_x$ into $Gd_2O_3$ and thus reduce the $CoO_x$ into Co. They realized the control of the magnetism of Co films by voltage-driven reversible oxidation and reduction, which enables manipulation of the magnetism.

Furthermore, in 2015, Bauer *et al.*[115] reported *in situ* observation of voltage-driven $O^{2-}$ migration in a Co/$GdO_x$ bilayer, where the interfacial magnetic anisotropy energy is quite susceptible to the oxygen state of the interface. Thus, the solid-state electro-chemical control of oxygen coordination by voltage could be utilized to modulate the magnetic anisotropy. The response speed of such a device can be increased by several orders of magnitude through changing temperature and gate voltage. The magneto-ionic switching time will be significantly shortened if the $O^{2-}$ diffusion barrier could be reduced, which can be realized by a decreasing thickness, or the morphology of the gate oxide and electrode is further optimized.

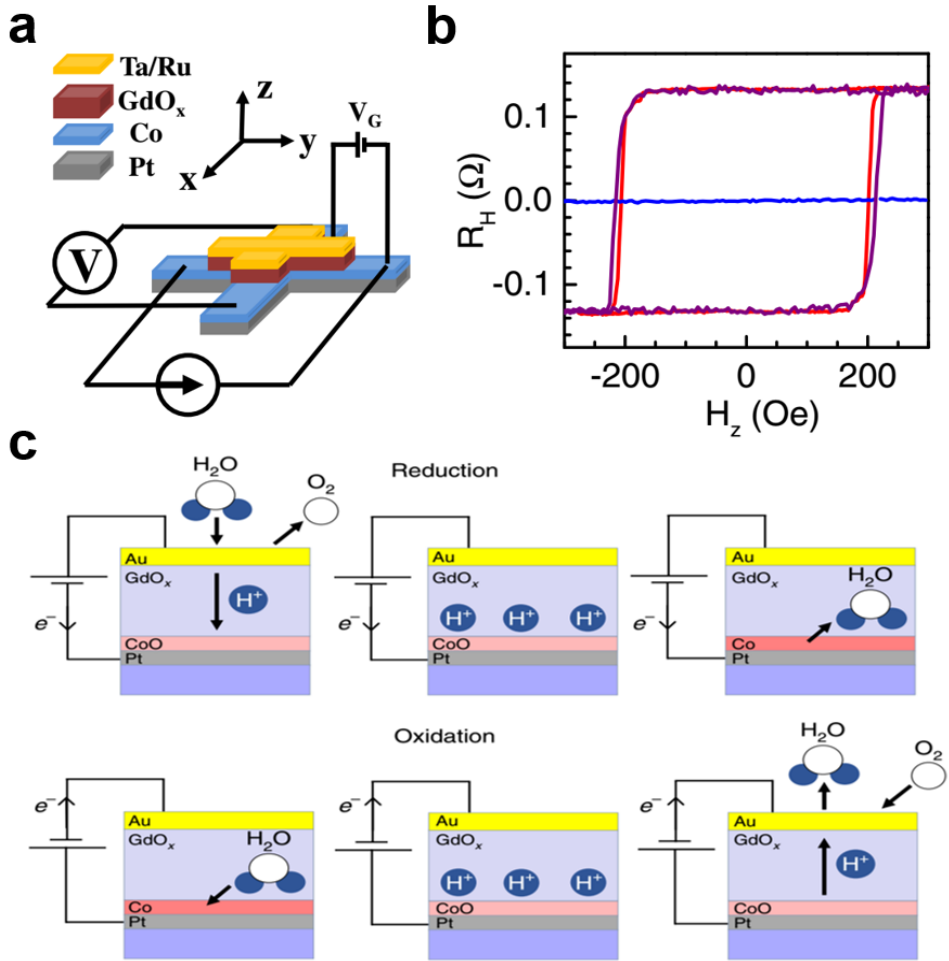

**Figure 17.** a) Schematic of a Ta/Ru/GdO$_x$/Co/Pt Hall bar structure and the transport measurement geometry. b) $R_H$ - $H_z$ curves of the sample after staying at 200 °C for 10 min (red line), after a -625 kV·cm$^{-1}$ electric field applied for 6 min (blue line) and a +625 kV·cm$^{-1}$ electric field applied for 13 min (purple line).[114] Copyright 2014, American Physical Society. c) Schematic of CoO reduction at a positive gate voltage and Co oxidation at a negative gate voltage.[116] Copyright 2019, Springer Nature-NPG.

Although O$^{2-}$ displacement in solid-state heterostructures can reversibly modulate magnetic anisotropy and magnetization, the problems such as irreversibility and device degradation exist because of the chemical and structural changes in ferromagnets at high temperatures that are required for these types of devices. In 2019, Tan *et al.*[116] proposed the non-destructive magnetic property manipulation with a small voltage, which utilizes H$_2$O hydrolysis as a solid-state proton pump. They applied a gate bias to a Pt/Co/GdO$_x$/Au heterostructure. During the reduction process under a positive gate voltage, H$_2$O is hydrolyzed at the anode while the oxygen evolution reaction produces H$^+$ and O$_2$. Then the H$^+$ migrates towards the bottom electrode, where the proton reacts with CoO to form Co and H$_2$O. During the oxidation under a negative gate voltage, Co is oxidized through the reverse process. Pt is the anode for Co oxidation and Au is the cathode for $H_2O$ recombination in this situation (Figure 17c). It is obvious that the proton rather than the $O^{2-}$ dominates the transport no matter the reduction or oxidation procedure. Afterward, magnetization in a Co film could be switched reversibly at room temperature through not only an interfacial $H^+$ inserting but also a hydrogen loading of an adjacent heavy-metal layer. The discovery lays a foundation on potentially high-speed, high-efficiency room-temperature ionic devices. The weakness of this approach could be that the modulation of each electronic state needs ~100 ms, and it may not be appropriate for fast and real-time device applications yet.

## 3. Outlook

### 3.1. Electric-field-controlled tunnel junctions utilizing metallic antiferromagnetic oxides

Overall, despite the rapid development of antiferromagnetic spintronics in recent years, the entire field is still at the infant stage compared with the mature ferromagnetic spintronic devices, such as the contemporary predominant information storage techniques – hard disks and data centers based on ferromagnetic thin films. From the memory device point of view, the signal read-out in antiferromagnetic devices is still difficult for practical applications as it mainly depends on the anisotropic magnetoresistance resulting from anisotropic density-of-states and relativistic spin-orbit coupling in antiferromagnetic materials, which is rather small at room temperature, typically on the order of 0.1-1%.[7,40,117] Therefore, significant long-term efforts are needed for developing room-temperature antiferromagnetic tunnel junction devices.

As learned from ferromagnetic tunnel junctions, interfaces are critical for achieving high anisotropic tunneling magnetoresistance, the exploration of metallic antiferromagnetic oxides could be promising considering the barrier layers in tunnel junctions are usually non-magnetic oxide insulators such as MgO and $Al_2O_3$. For example, the recent studies on rutile $RuO_2$ have illustrated that it is a conductive collinear antiferromagnet with a Néel temperature of ~400 K.[118,119] Our preliminary experimental results[120] found that the antiferromagnetic order also exists in its thin films, which are epitaxially grown on MgO and PMN-PT single-crystal substrates (**Figure 18**a-d), hence demonstrating the great epitaxial compatibility of $RuO_2$ with both ferroelectric oxides and tunnel barrier materials for actualizing full-oxide antiferromagnetic tunnel junctions. Remarkably, the electrical resistivity of $RuO_2$ thin films can be ultralow after optimizing growth temperature and oxygen partial pressure. As shown in Figure 18e, the lowest room-temperature electrical resistivity is obtained for $RuO_2$ thin films deposited at 550 °C and an oxygen pressure of $10^{-3}$ Torr. It reaches ~71 μΩ·cm, which is less than half of the resistivity for most of the antiferromagnetic intermetallic alloys such as CuMnAs,[14] MnPt,[40] $Mn_3Pt$,[20] $Mn_3Sn$,[12] and FeRh.[72] The low resistivity of this material could be favorable for reducing Joule heating during information read-out and spin-orbit torque information writing. Moreover, temperature-dependent resistance measurements indicate that the $RuO_2$ thin films fabricated at relatively high temperatures from 500-650 °C are all metallic (Figure 18f).

Therefore, it is promising to fabricate a $RuO_2$-based tunnel junction such as in terms of a Pt/MgO/$RuO_2$/PMN-PT heterostructure, where the antiferromagnetic spin axis of $RuO_2$ could be rotated by the electric-field induced piezoelectric strain from ferroelectric PMN-PT and thus the tunneling resistance of electrons from the top Pt electrode through the MgO insulating barrier to the bottom metallic $RuO_2$ could be mainly modulated by the electric field under zero magnetic field.

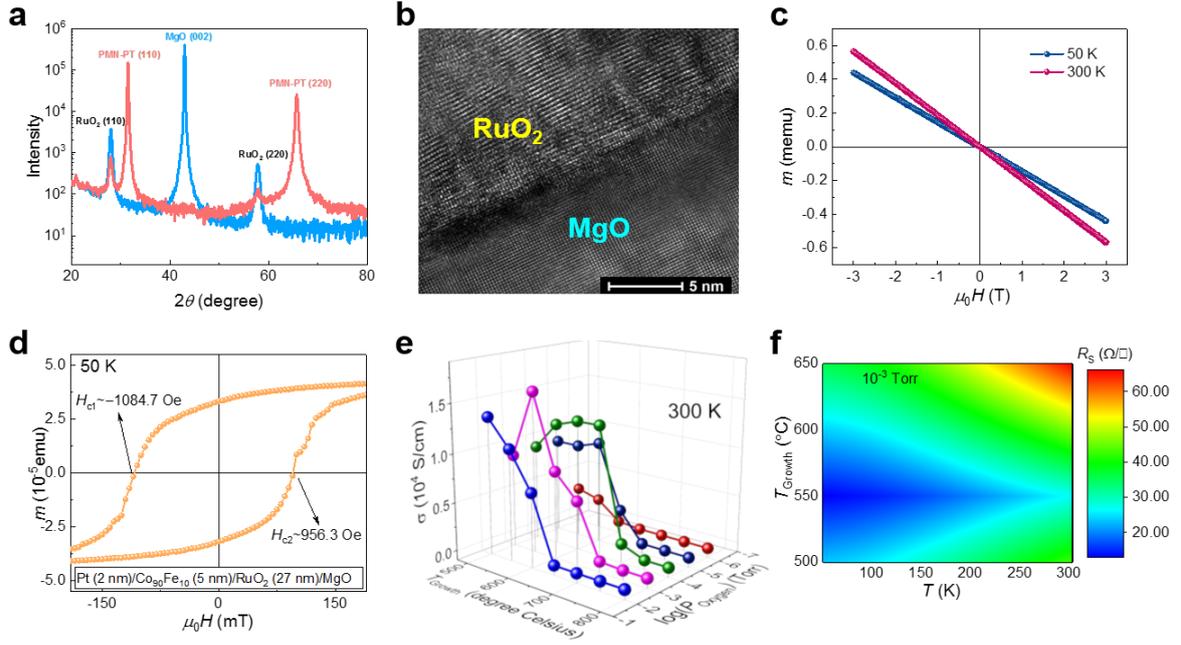

**Figure 18.** a) X-ray diffraction spectra of 27-nm-thick RuO$_2$/MgO(001) and RuO$_2$/PMN-PT(110) heterostructures. b) Cross-section transmission electron microscopy image of the RuO$_2$/MgO heterostructure. c) Magnetic moment versus magnetic field up to 3 T at 50 and 300 K for the RuO$_2$/MgO sample. d) Exchange bias of the RuO$_2$ film with a soft ferromagnetic Co$_{90}$Fe$_{10}$ layer at 50 K. e) Three-dimensional plot of the electrical conductivity of RuO$_2$ thin films grown MgO as a function of growth temperature and oxygen pressure. f) Contour plot of the sheet resistance of 27-nm-thick RuO$_2$ thin films fabricated at an oxygen pressure of 10$^{-3}$ Torr under different deposition and measurement temperatures.[120]

In addition to this long-term perspective on antiferromagnetic tunnel junctions for memory device applications, there could be another two points of view that may be of immediate interest to the antiferromagnetic spintronics society.

## 3.2. Electric-field-controlled antiferromagnetic spin logic devices

The traditional integrated circuits are based on complementary metal-oxide-semiconductor (CMOS) transistors, which have fallen into a bottleneck of development due to the quantum tunneling effect and the Boltzmann tyranny. Currently, the industrial progress on the size, voltage, and frequency scaling of the semiconductor chips has slowed down, and Moore's law is approaching failure. Therefore, a new concept of beyond-CMOS logic devices has been proposed to decrease the size and power consumption further and increase the response speed. Physical effects other than electron charges, *i.e.*, electron spins, magnons, photons, magnetic domain-walls, skyrmions, phase transition, and magnetoelectric coupling, have been exploited to perform logic computing. Correspondingly, various schemes of beyond-CMOS logic devices have been moved forward, such as spin transistors,[121,122] all-spin logic devices,[123] magnon spintronic devices,[124,125] all-optical transistors,[126,127] magnetic domain-wall logic devices,[128] skyrmion logic devices[129-131] and magnetoelectric spin-orbit logic devices.[132]

Currently, the spin logic devices have predominantly utilized ferromagnetic materials and relied on the switching of ferromagnetic moments to achieve logic functions. However, the logic devices based on antiferromagnetic spins have been rarely realized. Antiferromagnets have THz spin dynamics and insensitivity to stray magnetic fields. Accordingly, it could be inferred that the antiferromagnetic spin logic devices can have advantages of ultrafast THz switching speed, high packing density, and insensitivity to external perturbations of magnetic fields and bring a breakthrough in spin logic devices. In particular, using the electric field to control the antiferromagnetic spin logic devices could further reduce power consumption.

We now propose a new concept of the electric-field-controlled antiferromagnetic spin transistor-like device as schematized in **Figure 19**, which can turn on and off the spin current using a gate electric field by tuning the antiferromagnetic spin axis orientation. Specifically, this spin transistor could be composed of a Pt/$Cr_2O_3$/yttrium iron garnet (YIG)/PMN-PT multilayer structure. When the spin polarization of the spin current injected into the antiferromagnetic insulator $Cr_2O_3$[133] is parallel to the Néel vector of $Cr_2O_3$, the spin current can be transmitted through $Cr_2O_3$. In contrast, when the spin polarization of the spin current is perpendicular to the Néel vector, $Cr_2O_3$ cannot conduct the spin current. Consequently, we can control the transmitting of the spin current using an electric field, which can induce a piezoelectric strain in PMN-PT and thus manipulate the Néel vector of $Cr_2O_3$. Here, the spin current is generated by the spin accumulation in YIG through the spin Seeback effect and can be detected as a voltage signal through the inverse spin Hall effect in the Pt layer.

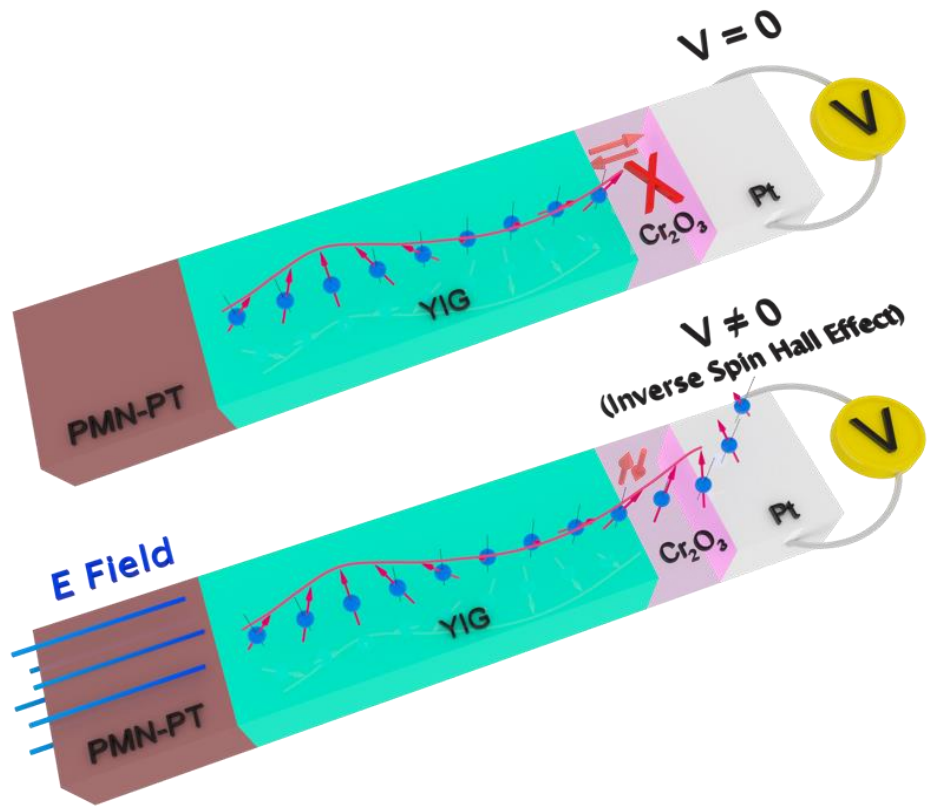

**Figure 19.** Illustration of the electric-field-controlled antiferromagnetic spin transistor.

**3.3. Electric-field-controlled antiferromagnetic artificial neuron devices**

The human brain transmits and processes information based on neurons and synapses, which is substantially different from conventional CMOS logic devices. The artificial neurons simulating the operation and structure of biological neurons have a new architecture and the advantages of ultrafast speed, low energy consumption, and error tolerance relative to conventional information computing devices. A characteristic of the operation of neurons is the spike signal, which is called action potential in biology. It is worth noting that the antiferromagnetic materials have the THz response speed, which can generate picosecond-scale spike signals, and are suitable for artificial neuron devices.

In 2018, Khymyn *et al.* theoretically proposed an artificial neuron device based on a current-driven antiferromagnetic auto-oscillator comprised of an antiferromagnet/normal metal heterostructure.[134] When the applied current density increases to exceed the threshold, the artificial neuron device can excite a single sharp spike or a train of periodic spikes owing to the non-uniform rotation of the Néel vector in the antiferromagnet driven by the current-induced spin-transfer torque. The duration of the spike is 2.4 ps corresponding to the THz oscillation of antiferromagnetic moments. With further increase of the applied current density, a discrete group of spikes can be caused similar to the behavior of bursting signals in the natural neurons. The concatenability is an essential characteristic of logic devices, and this scheme has also achieved the connection of two artificial neurons through a tunable amplifier resembling a synapse (**Figure 20**).

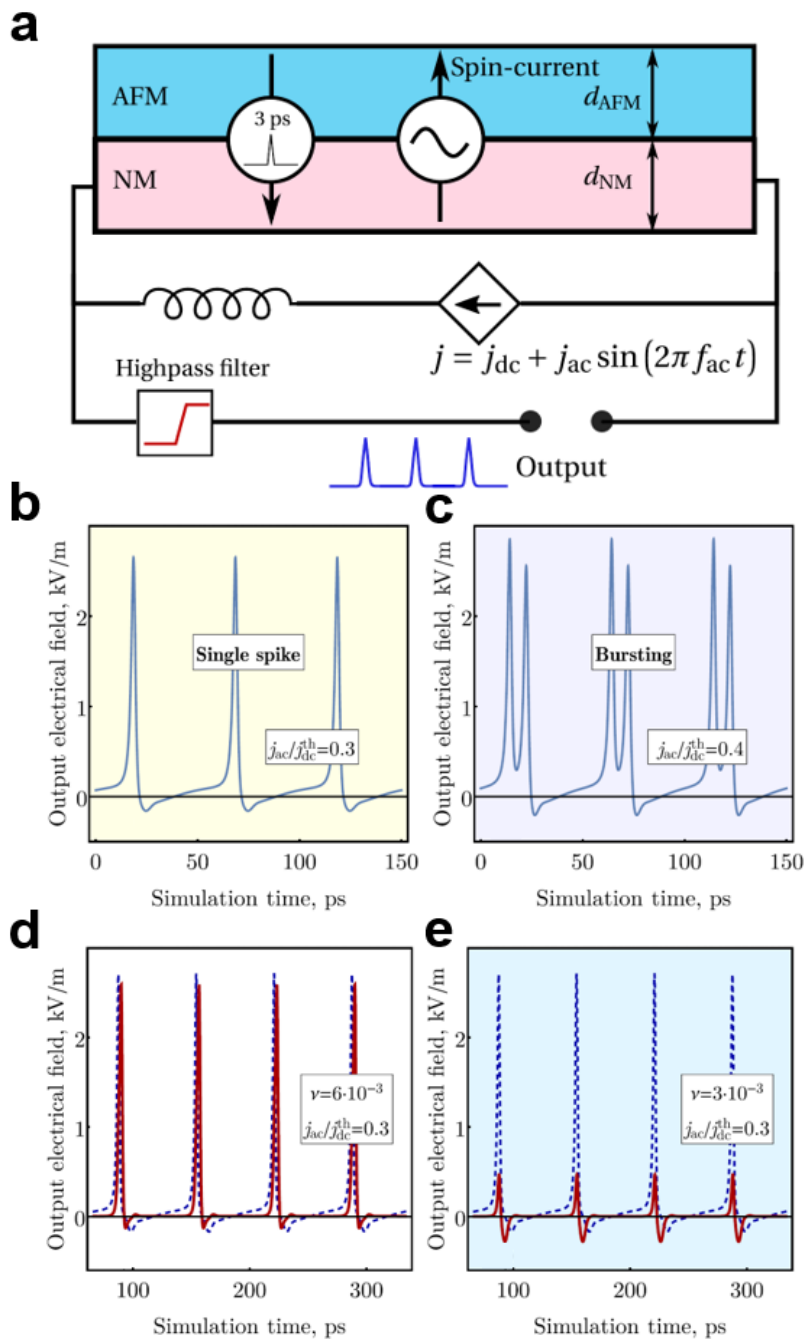

**Figure 20.** a) Sketch of the spikes generator of an antiferromagnetic artificial neuron device. Generation of b) a single spike and c) a discrete group of spikes. The spikes of d) the anterior and e) the posterior neurons in a neural network.[134] Copyright 2018, Springer Nature-NPG.

Soon after that, the same research team further proposed that the antiferromagnetic neurons mentioned above can, in principle, perform the basal logic functions.[135] When the applied current density is relatively small, and only both of two input spike signals are superimposed to exceed the threshold, the output spike signal can be received. This operation is similar to the AND logic gate. In the same manner, when either any one of the input signals is relatively large and can exceed the threshold to excite an output signal, the OR logic operation is performed. The majority gate and Q-gate can also be achieved in the antiferromagnetic artificial neurons, and even the full-adder comprising one M-gate and two Q-gates (**Figure 21**). These theoretical research works suggest that the antiferromagnetic artificial neurons could be rather promising for ultrafast information transmitting and processing neuromorphic system applications, which are pending for experimental test and verification.

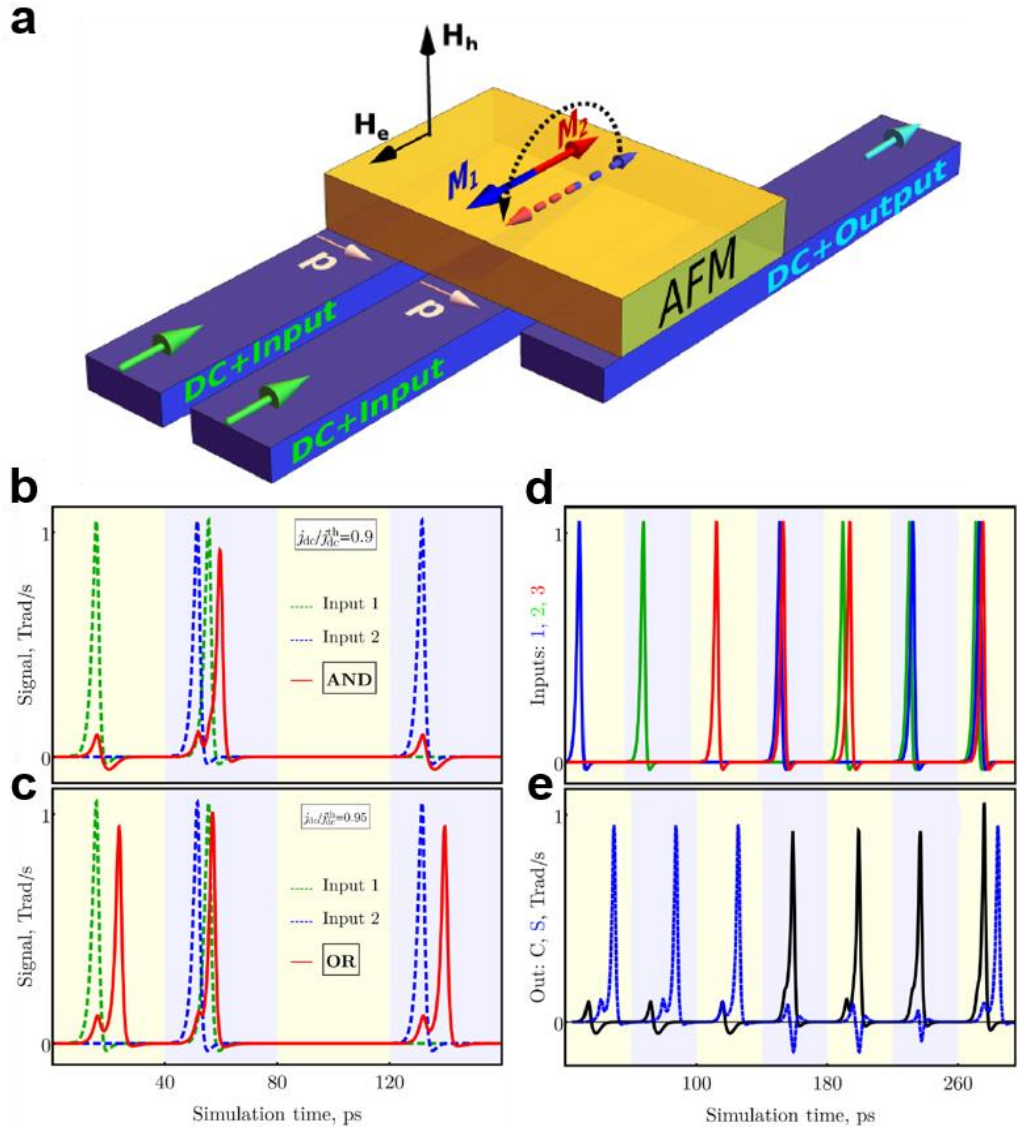

**Figure 21.** a) Schematic of an antiferromagnetic artificial neuron device. The processing of signals in b) the OR logic gate and c) the AND logic gate based on the antiferromagnetic artificial neuron devices. d) The input and e) the output signals in a full-adder based on the antiferromagnetic artificial neuron devices.[135] Copyright 2018, American Institute of Physics.

The antiferromagnetic artificial neurons driven by electrical currents cause energy waste due to the Joule heating effect, which degrades their energy efficiency. Because the electric-field control of antiferromagnetic devices can avoid the Joule heating, electric fields are more appropriate than currents to switch the antiferromagnetic moments. The electric-field-controlled antiferromagnetic artificial neuron devices could have an apparent advantage of ultralow energy consumption in accordance with biologic neural systems.

**Acknowledgements**
Zhiqi Liu acknowledges financial support from the National Natural Science Foundation of China (NSFC Grant No. 51822101, 51861135104, 51771009, & 11704018)